\documentclass{aa}  

\usepackage{amsmath}
\usepackage{makecell}
\usepackage{graphicx}
\usepackage{upgreek}
\usepackage{epstopdf}
\usepackage{txfonts}
\usepackage{colortbl} 
\usepackage[table,xcdraw]{xcolor}
\begin{document} 

\renewcommand{\textbf}[1]{#1}
\renewcommand{\mathbf}[1]{#1}
\renewcommand{\bfseries}[1]{#1}
\title{The influence of stellar activity on detecting Earth-like planets via nulling interferometry}



\author{Rui-Si Zhou\inst{1,2}, Hui-Gen Liu\inst{1,2}\thanks{Corresponding author: \email{huigen@nju.edu.cn}}, Li-Yong Zhou\inst{1,2}}

\institute{School of Astronomy and Space Science, Nanjing University, Nanjing, 210023, People’s Republic of China
         \and
         Key Laboratory of Modern Astronomy and Astrophysics, Ministry of Education, Nanjing, 210023, People’s Republic of China}

   \date{Received September 15, 1996; accepted March 16, 1997}

 
  \abstract
  {To investigate the influence of stellar activities on the detection of Earth-like planets via the \textbf{nulling interferometer}, we aim to estimate the signal-to-noise ratio (S/N) values of Earth-like planets around typical dwarf stars G and M, with different stellar activities. We also study the fitted accuracy of the planetary locations with different S/N values, which is crucial to evaluating the possibility of a planet in the habitable zone (HZ).}
   {The direct imaging of Earth-like planets in solar neighbors is challenging. Both transit and radial velocity (RV) methods suffer from noise due to stellar activity. Here, we focus on the  \textbf {differential nulling interferometer with an X configuration} and examine whether stellar activities are likely to influence the detection of planets in the  HZ and the location accuracy. We aim to provide a basis for selecting target stars according to their specific activity levels for future interferometer observatories in space.}
   {By choosing a typical configuration of an \textbf{X array interferometer}, we used theoretical formulas to calculate the intrinsic Poisson noise and the noise of stellar activities.\textbf{ Assuming a fixed array with no rotation and} ignoring other systematic and astrophysical noises, we considered a single active region on a stellar disk, including both spots and flares with different parameters, for instance, the position, size, and temperature of the active regions. Then we simulated the S/N of Earth-like planets in HZ around G dwarf stars (solar-like) and M dwarfs (Proxima-like), with different stellar activities in the mid-infrared (MIR) band (7-12 $\mu$ m). Furthermore, we used analytic and numerical ways to investigate the influence on the determination of the planet's location, based on the deduced S/N caused by the stellar activity.}
   {The noise attributed to stellar activity has much less of an influence than the TV and RV when attempting to detect Earth-like planets around both G and M dwarfs.  Stellar activity may hardly influence the detection of Earth-like planets around G dwarf stars,\textbf{ with a S/N exceeding 10 for both flares and spots even when the active region reaches three times Jupiter’s radius}. \textbf{ However, detecting Earth-like planets around M dwarfs, which are usually more active, can be significantly hindered. If the stellar activity is violent enough, the S/N can drop below 5 for both flares and spots when the active region is as large as three times Jupiter’s radius.} We also analyzed the uncertainty of the planet's location due to the deduced S/N. Consequently, we have determined the possibility of mistaking a planet in the HZ as being outside the HZ based on an erroneous S/N measurement.}
   {\textbf\textbf{ Since the nulling interferometer largely suppresses the stellar signal, the impact of stellar activities, such as spots and flares, on the detection of Earth-like planets is significantly reduced. However, for some extremely active M stars, such activities can make the detection of planets obscure or lead to greater uncertainties regarding their location. Selecting quiescent target stars or monitoring the light curves of stars would be a helpful way to get rid of contaminates associated with violent stellar activities.}}

\titlerunning{Does the stellar act influence the detection of earth-like planet in HZ via interferometry}
\authorrunning{Ruisi-Zhou et al.}
\maketitle
%

\section{Introduction}
\label{sec:1}
\begin{figure}[h]
    \centering
    \includegraphics[width=1.04\linewidth]{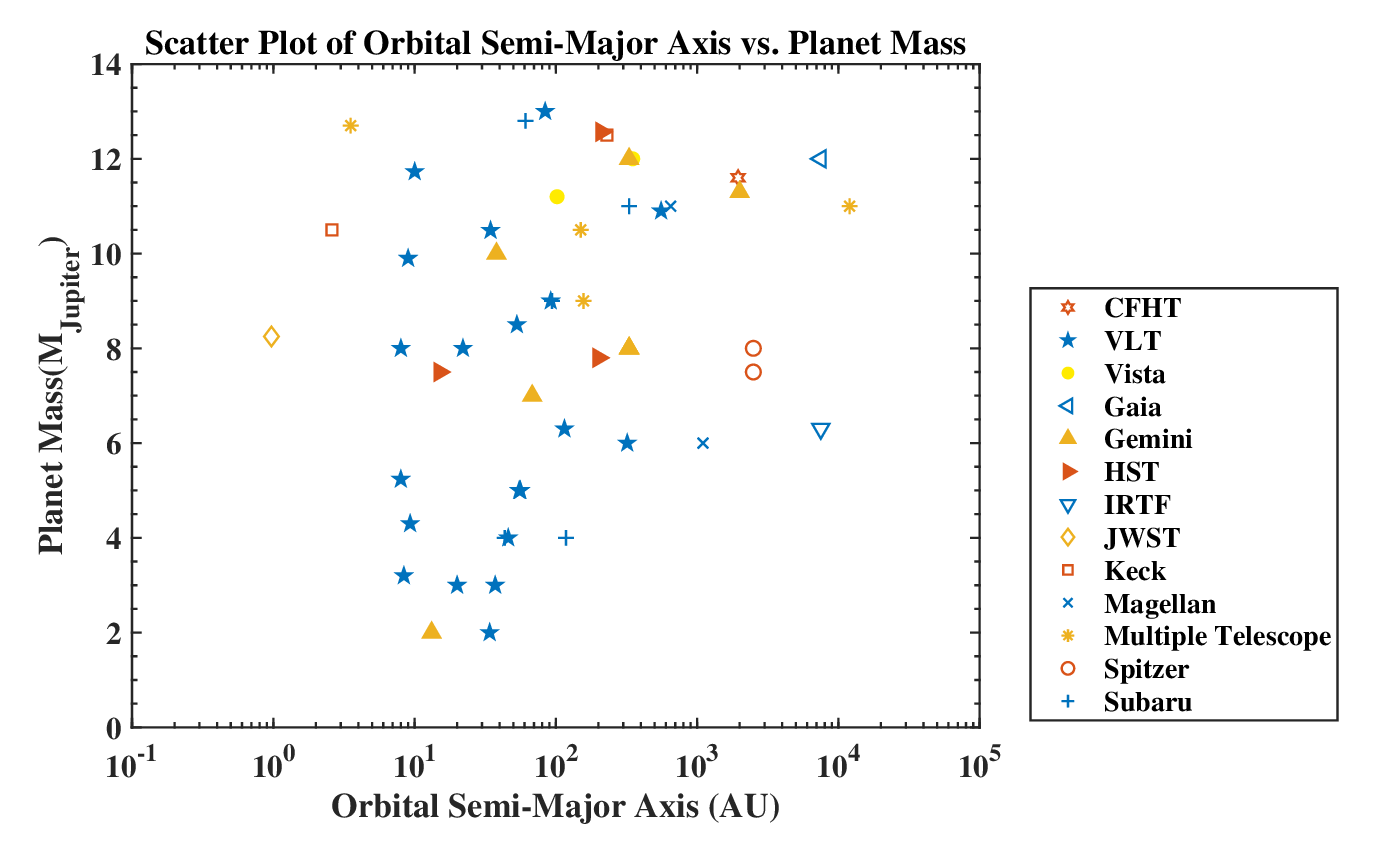}
    \caption{Masses and orbital semi-major axes of exoplanets discovered by direct imaging methods. Different symbols represent observations from different telescopes, as indicated in the legend.}
    \label{fig1}
\end{figure}

Currently, the most popular methods for detecting exoplanets are the transit (RV) and radial velocity (RV). Thus far, more than 93\% of exoplanets have been detected via these two methods, including some terrestrial planets in the habitable zone (HZ) around nearby M-dwarfs (e.g., Proxima b, Trappist-1e, and others ). To characterize such planets, direct imaging is an essential way of obtaining signals from the surface of exoplanets. Coronagraphs are also utilized to block stellar photons and reduce the contrast between the star and the planet. The coronagraph technique is widely used on many ground-based telescopes, for instance, KPIC on KecK \citep{a11}, SPHERE on VLT \citep{a12}, and GPI on Gemini \citep{a13} to image exoplanets. \textbf{\citep{R0} predicts the detection of nearby rocky exoplanets from the METIS on ELT}. The coronagraph indicates preferential detections of young massive gas giant planets far away from the host star. As shown in Fig. \ref{fig1}, there are 53 giant planets with masses of $\leq 13 M_{\rm J}$ ($\sim1\%$ of all detected planets)  detected via the imaging method. All of them are more massive than Jupiter and beyond 1 AU. The direct acquisition of the thermal radiation from planets and the spectroscopic analysis of molecules in their atmospheres are the basis for studying the chemical components of these giant planets and indicate their formation mechanisms \citep{1,2}.

However, to image Earth-size planets in the  HZ (hereafter, these objects are referred to as Earth-like planets) around stars is challenging, even for the nearest planet, Proxima Centauri b. The biggest 
challenge lies in the extreme contrast and spatial resolution, namely, the faint terrestrial planetary signals are usually hidden in the signals of the bright host stars. Currently, no constructed telescope can image terrestrial exoplanets in HZ.
Even with the powerful JWST equipped with a coronagraph in NIRCam, the detection capability mainly depends on the inner working angle (IWA) and the achieved contrast, which ranges between 0.13'' and 0.88'' for the IWA, with a contrast of $10^{-3}$ to $10^{-2}$. It is far from the requirements to image Earth-like planets,  typically in the case of a Sun-Earth system at 10 parsecs. In the future, the Habitable World Observatory (HWO) will provide a 10 m space telescope equipped with a more powerful coronagraph with a contrast of $10^{-10}$, aiming to image the nearby rocky planets in HZ.\citep{3}

The IWA of coronagraph is limited by the Rayleigh criterion, which depends on both the wavelength and the diameter of the telescope. To achieve a higher resolution, the concept of a distributed synthetic aperture is proposed. In particular, the configuration of the telescope array, using interference to reduce the stellar light intensity and deduce the planet-star contrast, is another path to imaging Earth-like planets, namely via nulling interferometry. Compared to a coronagraph, nulling interferometry can achieve higher spatial resolution by stretching the baseline, but it is less efficient when we are looking to obtain the images of extended sources, due to the need to adjust baselines in the sampling process to cover the u-v plane. 

In this paper, we consider the detection of planets based on nulling interferometry techniques. The approach was first proposed \citep{4} with two telescopes. Interferometers with more telescopes are also proposed, which use a linear operator to describe the combination of the light from different telescopes, with phase compensations in each output. By solving the linear equations, the interferometer setups can be optimized to eliminate the effects of optical path differences between apertures \citep{5}.\textbf{The X array configuration uses a dual chopped Bracewell to eliminate spatial symmetric noise via differential outputs, while the \textbf{differential nuller} prefers a non-redundant array configuration. Dual chopped Bracewell has been used on Keck Interferometer Nuller \citep{R1} or is currently planned on NOTT for VLTI \citep{R2}}. Previous conception Mission Darwin and TPF-1 proposed the Space Interferometry Programme\citep{6}, which is meant to use distributed configuration to achieve nulling interferometer. However, these two missions did not step into the implementation phase. As an inheritor, the LIFE project is also conducting research on space interferometry \citep{7} and comparing the performance of various interferometer configurations \citep{8}. The effect of various noises during the observations is considered, such as local zodiacal light,  exozodiacal light, and other effects \citep{9}. We chose a basic configuration containing four telescopes to analyze the detection of Earth-like planets.

        
Exoplanet detections based on traditional methods are affected by stellar activity. For example, both the transit and RV measurements suffer from stellar activity noises, especially when we are attempting to detect terrestrial planets. In the case of transit, the stellar activity reduces the relative photometric accuracy (several tens to a hundred ppm for quiet stars and hundreds of ppm for active stars), which can hide the transit signals of small planets \citep{10}. Furthermore, the characterization of exoplanets can be affected by stellar activity. The M8V star TRAPPIST-1, which has seven Earth-sized planets \citep{11}, is a promising target for exoplanet characterization. Optical observations from the Kepler/K2 Campaign of TRAPPIST-1 indicate a 3.3 day period with clear rotational modulation. However, this period may not represent the star's rotation but, rather, the varying timescale of active regions on the star \citep{12}.

For RV measurements, active regions on the stellar surface (for instance, spots or flares) can lead to variations in the stellar spectral lines, which cause RV shifts. The typical RV jitter is $\sim$ 1-2 m/s, and even up to 10 m/s for active stars \citep{13}. For instance, the radial velocity variations of HD 166435 were thought to be caused by a companion star, but further research by \citet{14} revealed that it was actually due to surface activity on the host star, such as spots. In this paper, we aim to examine whether the effect of stellar activity also disrupts the detection of Earth-like planets when measurements are performed using a nulling interferometer.

In the photosphere,  stellar activity is usually manifested as dark spots or bright flares. We focus on these two phenomena to check how they raise additional noises in a nulling interferometer. In this paper, we estimate the signal-to-noise ratio (S/N) of interferometric imaging detections of exoplanets in wide mid-infrared bands \textbf{(7-12 $ \upmu \rm m$)} and study the affects of parametric stellar activities. Additionally, we investigate the decrease in the S/N and analyze the influence on the position uncertainty of potentially habitable planets, which is crucial for determining whether the planets are located in the HZ or not.

The paper is organized as follows. In Section \ref{sec:2}, we introduce the spectra model of stellar and planetary, as well as the spectral models used for stellar activity (spots, flares), along with our choice for the optimized configuration of the nulling interferometer. In Section \ref{sec:3}, we analyze the influence of the position, size, and temperature of the active region on the detection of planets around solar-like stars. To compare the ability to detect HZ Earth-like planets around solar-like stars and red dwarfs, we also compare the effects of stellar activity on M dwarfs in Section \ref{sec:4}. Furthermore, we analyze how the decreasing S/N due to stellar activity affects the determination of planetary locations in Section \ref{sec:5}. Finally, we summarize and discuss our conclusions and we also expand on other noise effects that may impact the detection results in Section \ref{sec:6} and \ref{sec:7}, respectively.

\section{Methods}
\label{sec:2}
\begin{figure}[h]
    \centering
    \includegraphics[width=1\linewidth]{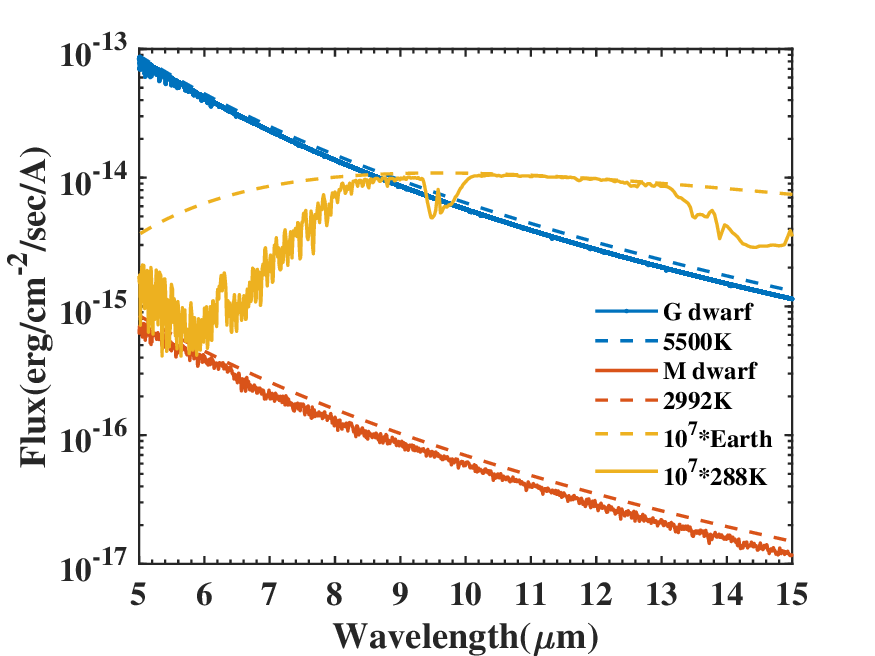}
   \caption{\textbf{Spectrum of stars and planets. The solid line represents the emission spectrum of an Earth-like planet generated by MODTRAN(black), the simulated spectrum of a solar-like G dwarf star (blue), and a Proxima-like M dwarf star (red), while the dashed line shows the spectrum calculated using the Planck formula at 288K(black), 5500 K(blue), and 2992 K(red), respectively.}}
    \label{fig2}
\end{figure}

To study the noise generated by stellar activity during exoplanet observations, we found it is necessary to model the spectra of quiescent stars and planets in the observational mid-infrared (MIR) band. We used the Phoenix model \citep{17,16} to generate the quiescent spectra of solar-like stars. To obtain the infrared radiation of Earth-like planets with a surface temperature of 288 K, we used the 1976 standard atmospheric model to simulate the infrared radiation of the planets via Modtran \citep{24}, including the absorption characteristics of important molecules such as water and carbon dioxide. As a typical case, we set an Earth-like planet at 1 AU around a solar-like star, with a distance of 10 pc. The spectrum of both star and planet are demonstrated in Fig. \ref{fig2}. 

Compared with G dwarfs, M dwarfs dominate the stellar population in solar neighbors \citep{23,a8}. Both transit and RV methods are more likely to detect Earth-like planets around M dwarfs rather than G dwarfs. Thus, we also considered the case of detecting planets in HZ around M dwarfs. Stellar parameters similar to Proxima Centauri are adopted to represent a typical M dwarf (see Table \ref{table1}). The spectrum is also obtained from the Phoenix model (see Fig. \ref{fig2}). We assume the Spectrum of Earth-like planets around M and G dwarfs are the same; that is, they have the same surface temperature of 288 K and Earth-like atmosphere. We note that considering chemical equilibrium in the atmosphere undergoing different radiations, the planet around different stars may have different components and lead to different emission spectrums \citep{a9,a10}. 

As shown in Fig. \ref{fig2}, the simulated stellar spectrum in 7-12 $\upmu$m  are $\sim$10\% less than the blackbody spectrum for both G and M dwarfs. Furthermore, the spectrum of the M dwarf is not as smooth as the G dwarf stars in Phoenix models. Although the blackbody spectrum can approximately represent the infrared spectrum of stars, the emission spectrum of the planet deviates from the blackbody in a more obvious way. For example, there is much less emission around 7 $\upmu$m due to water absorption and additional emission from the atmosphere around 10-12 $\upmu$m, compared with the blackbody radiation of 288 K. 

\begin{table}[h]
\caption{The stellar and planetary parameters for M dwarf and G dwarf star systems and the interferometer parameters.}
\setlength\tabcolsep{3pt}
\renewcommand\arraystretch{1.5}
\begin{tabular}{c|c|c|c}
\hline
\textbf{Symbols} & \textbf{M dwarf}\tablefootmark{a}& \textbf{G dwarf} & \textbf{Description} \\ \hline
\multicolumn{4}{c}{\bf{Star-planet parameters}} \\
\hline
d                 & 10 pc        & 10 pc       & Target distance                    \\ 
$\theta_p$         & 4.03 mas      & 100 mas       & Planet-star separation                    \\ 
$R_p$ & 1 $R_{\oplus}$ & 1 $R_{\oplus}$ & Planet radius                    \\
$T_{p} $ & 288 K & 288 K         & Planet surface temperature                    \\ 
$R_s$                 & 0.15$R_{\odot}$     & 1.0 $R_{\odot}$         & Stellar radius      \\ 
$\theta_s$         & 0.069 mas        & 0.46 mas       & angular radius of star                \\
$T_s $              & 2992 K         & 5773 K      & Stellar effective temperature          \\
\hline
\multicolumn{4}{c}{\bf{Interferometer parameters (7-12 $\upmu$m)}} \\
\hline
$D$ &4 m&4 m&Telescope diameter        \\ 
$L$  &  131m   & 5.5m  & Half of the baseline \\
$L_{10 \upmu m}$  & 140.40 m     & 5.56 m    & Calculated $L$ at 10 $\upmu$m \tablefootmark{b}  \\
$t_{exp}$&5000 s &20000 s&\textbf{Total exposure time}\\
$\eta_{QE}$ &\textbf{0.5}&\textbf{0.5}& Quantum efficiency \\
$\eta_{t}$ &\textbf{0.02}&\textbf{0.02}&Instrument throughput\tablefootmark{c}\\
\hline
\end{tabular}

\tablefoot{
\tablefoottext{a}{The parameters of Proxima are adopted to represent M dwarfs \citep{a1}.\\} 
\tablefoottext{b}{$L_{10 \upmu m}$ represents the half length of baseline with a maximum planet signals at 10 $\upmu $m, while $L$ represents the half length of baseline with maximum planet signals in the band of 7-12 $\upmu$m, which depends on the spectrum of planet, and slightly smaller than $L_{10 \upmu \rm m}$. $L$ is adopted in our simulations.\\}
\tablefoottext{c}{\textbf{The total efficiency of the telescope is 0.01, determined by the product of the quantum efficiency and the instrument throughput. The integrated flux is proportional to the product of the exposure time and the total efficiency of the telescope.}}}


\label{table1}
\end{table}

After we generated the spectra of the quiescent star and planet, we also need the spectrum of the active regions to analyze the influence of stellar activity. Usually, stellar activities are illustrated as either spots or flares. In this paper, we present our modeling of their spectrum in MIR via blackbody radiation, with different temperatures. For a typical solar-like star, the temperature of spots ranges from 1000-5000 K, while the temperature of the flare is from 9000 to 10,000 K \citep{18}. For typical M dwarfs, spot temperatures range from 500 to 2000 K, while flare temperatures can reach  20,000 to 30,000 K\citep{19,20,21}. Therefore, we set the flare temperature for M dwarfs in the range between 5000 and 30,000 K. The area of the active region of the spot or flare is also crucial for determining the strength of stellar activity and we set two typical sizes of active regions, namely, Earth-sized and Jupiter-sized \citep{22}. More details and settings can be found in  Sects. \ref{sec:3} and \ref{sec:4}.

To determine the configuration of nulling interferometry,\textbf{ we adopt the X-array configuration interferometer with four telescopes} \textbf{\citep{ R4}}. The X-array of the telescopes has two sets of baselines in two orthogonal directions, where the ensemble matrix $\bf{\rm U}$ is set up the same as Equation (3) in \citep{9}. The ratio of the baselines in two orthogonal directions $q$ affects how well the \textbf{instability noise} can be corrected. According to \citep{15}, it is better to remove \textbf{instability noise} if the ratio is set as 1:6 rather than 1:1. \textbf{This conclusion is based on a fixed wavelength of $10 \upmu \text{m}$, but we here consider a wavelength range from $7 \upmu \text{m}$ to $12 \upmu \text{m}$.} Thus, it is necessary to optimize the ratio $q$ for the \textbf{nulling interferometer}.

\begin{figure}[h]
    \centering
    \includegraphics[width=0.5\linewidth]{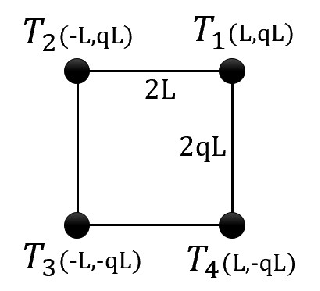} \\ 
    \includegraphics[width=1\linewidth]{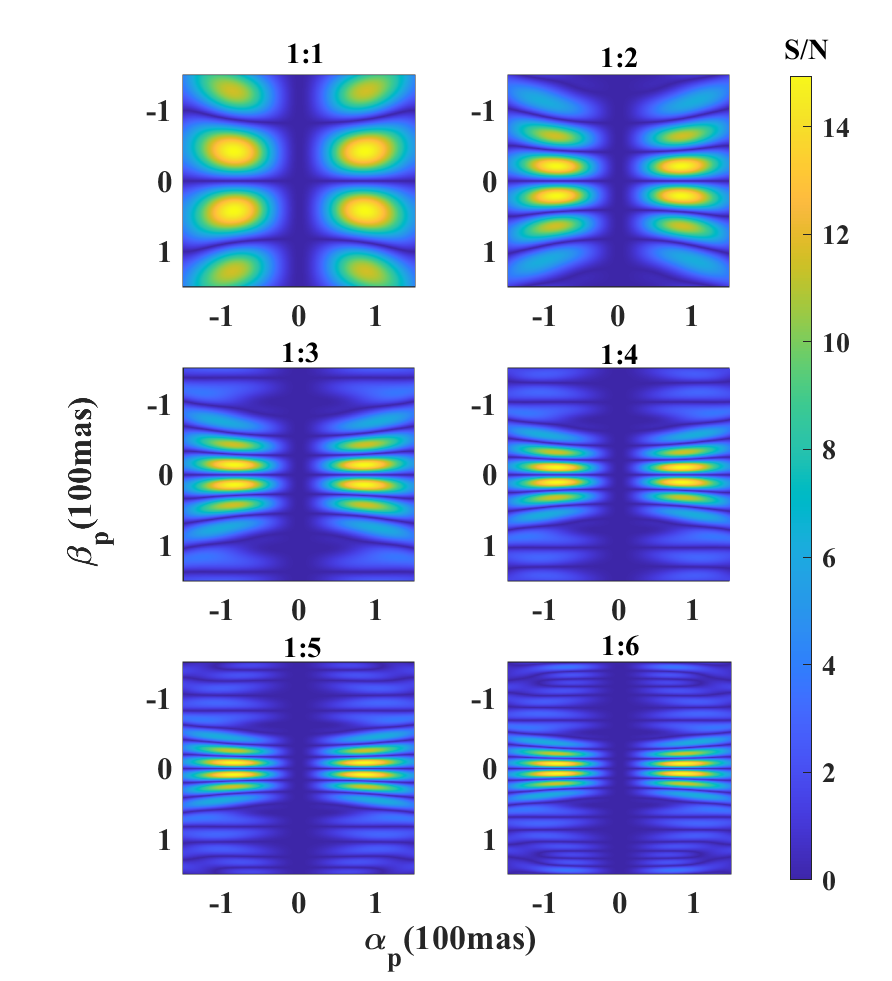} 
      \caption{Top sketching map shows the configuration adopted here and hereafter (corresponding to the transfer matrix in Eq. (6) from \citealp{8}). The color maps show the distribution of the S/N for an interferometer detecting an Earth-like planet around a G-dwarf (Sun-like) star at 10 pc. The panels from top to bottom show the S/N distribution for different configurations (X configurations) with baseline length ratios from 1:1 to 1:6. The S/N threshold is set at 10 to define the effective observation region where S/N $\geq$ 10.When the baseline length is \(1:1\), the maximum S/N (shown in the top-left quadrant) is achieved near \(\alpha:\beta=2:1\); specifically, at \((89, 44)\) mas.}

    \label{fig3}
\end{figure}

 \textbf{ The X-array outputs with a fixed baseline limit the ability to search for Earth-like planets in the HZ. A specific baseline and a non-rotating telescope configuration result in a high S/N only for planets located in certain regions. As illustrated in Fig. \ref{fig3}, with a baseline length of 11 m, only some areas achieve an S/N greater than 10.} To cover the whole HZ around G dwarfs, we need to rotate and adjust the length of the baseline. Thus, we can use the total time cost to cover the whole HZ to optimize configurations with different values for $q$. We define detection efficiency as the fraction of areas that satisfy the S/N threshold (i.e., the S/N of Earth-like planets is S/N$\ge$10) in the HZ. To minimize the time cost, we should choose a baseline ratio with maximum detection efficiency. To estimate the S/N, we only consider the Poisson noise in the wide band of 7-12 $\upmu$m and fixed a uniform exposure time of 20000 sec. Assuming a typical planetary system around a G dwarf (the parameters are shown in Table\textbf{\ref{table1}}), we display the efficiency parameters in Table \textbf{\ref{table2}}. The 1:1 configuration with equal baselines (11 m) has the highest observing efficiency of 1.69\%, while the 1:6 configuration has a much lower efficiency of 0.28\%. It is also obvious, as shown in Fig. \ref{fig3}, that the bright region of the 1:1 configuration is larger than that of the 1:6 configuration. Thus, in the following sections, we chose the 1:1 X-array as the configuration of the {nulling interferometer}. 

\begin{table}[h]
\caption{Maximum S/N and detection efficiency for telescopes with different baseline length ratios for the X configuration. }
\renewcommand\arraystretch{2.5}
\resizebox{\linewidth}{!}{
\begin{tabular}{c|c|c|c|c|c|c}
\hline
\textbf{X-Array}    & 1:1   & 1:2   & 1:3   & 1:4   & 1:5   & 1:6   \\ \hline
\textbf{\makecell{Maximum\\$\rm S/N$}}        & \textbf{15.55} & \textbf{15.55} &\textbf{15.55} & \textbf{15.55} & \textbf{15.55} & \textbf{15.55} \\ \hline
\textbf{\makecell[c]{Detection\\Efficiency }} & {\makecell{\textbf{10.18}\%}}  & 
{\makecell{\textbf{5.09}\%}}& 
{\makecell{\textbf{3.39}\%}}   &  {\makecell{\textbf{2.55}\%}}  &  {\makecell{\textbf{2.07}\%}}  &  {\makecell{\textbf{1.70}\%}}    \\ \hline
\end{tabular}}
\label{table2}
{\bfseries\tablefoot{The short baseline length is 11 meters, and the long baseline length is \( c \times 11 \) meters, where \( c \) is the baseline length ratio.}}
\end{table}

 When searching for Earth-like planets around stars via a nulling interferometer, both systematic noises and astrophysical noises affect the S/N of planets. For example, zodiacal light and exozodiacal light have been analyzed by TPF and LIFE \citep{8}, respectively. The main purpose of the present paper is to analyze the noise induced by stellar activity and its impact on exoplanet detection. Thus, we ignored all systematic noise and other astrophysical noise, such as background noise due to zodiacal light. To simplify the stellar parameters, in our simulations, we consider the star as a uniform centrosymmetric disk. Ignoring the pointing and other systematic errors, the symmetry of the stellar signals received by the interferometer will be guaranteed, allowing the \textbf{differential}  output to reach a null depth of $10^{-6}-10^{-7}$ for solar-like stars, which is dominated by Poisson noise. However, when stellar activity exists, the position of spots or flares is randomly distributed on the stellar surface, which can destroy the central symmetry and affect the \textbf{null depth} of the \textbf{X-array} nulling interferometry. 
 
 As shown in Eq. \ref{eq:1}, $N_{s}$ denotes the total output of stellar photons after \textbf{differential nulling} at the wavelength $\lambda$; then,  $\theta_{s}$ is the angular diameter of a star, a point in the sky at position $(\alpha,\beta)$ with a flux of $I_{s}(\alpha,\beta)$ in unit area, and $L$ is the half-length of the baseline. When the flux of a stellar disk is distributed uniformly and is central symmetric, $N_{s}$ should be zero if there is no Poisson noise. Equation \ref{eq:1} is directly derived from the Equation 10 in \citep{9}.\textbf{We separately calculated the Poisson noise for the X array on the outputs  3 and 4. According to the error propagation, the total Poisson noise of the differential output between outputs 3 and 4 is represented in Eq. \ref{eq:2}, where the $N_{s, total}$ is the total stellar photons before nulling. We note that here we assume the star as a uniform disk and the stellar angular diameter, $\theta_{s}$, is small enough to adopt $\sin{2\pi L \theta_{s}/\lambda} \approx 2\pi L \theta_{s}/\lambda$ approximately.} Compared with the Poisson noise of a star (i.e., $\sqrt{N_{s, total}}$), there is a small factor of $\frac{\sqrt{2}\pi L \theta_{s}}{\lambda}$, which is around 0.01 for G dwarf at 10 $\upmu$m, to suppress the stellar noise. Furthermore, Eq. \ref{eq:2} also hints at the area with larger $\alpha$ contributing more Poisson noise than other areas. As can be expected, the location and intensity of the stellar activity are important parameters that affect stellar leakage. Thus, we chose to investigate these factors in the detection of Earth-like planets around G and M dwarfs.
 \begin{equation}
  N_{s}=\int^{\theta_{s}}_{-\theta_{s}}
  \int^{\sqrt{\theta^2_{s}-\beta^2}}_{-\sqrt{\theta^2_{s}-\beta^2}} {4I_{s}(\alpha,\beta)}\sin^2(\frac{2\pi L \alpha}{\lambda})\sin(\frac{4\pi \textbf{q}L \beta}{\lambda})d\alpha d\beta   
  \label{eq:1}
,\end{equation}

\begin{equation}
\begin{aligned}
N_{\text{poisson}}&=\mathbf{\left(\int^{\theta_{s}}_{-\theta_{s}}
  \int^{\sqrt{\theta^2_{s}-\beta^2}}_{-\sqrt{\theta^2_{s}-\beta^2}} I_{s}(\alpha, \beta) \left(T_3+T_4\right) d\alpha \, d\beta \right)^{1/2}}\\
   &=\left(\int^{\theta_{s}}_{-\theta_{s}}
  \int^{\sqrt{\theta^2_{s}-\beta^2}}_{-\sqrt{\theta^2_{s}-\beta^2}} 4I_{s}(\alpha, \beta) \sin^2\left(\frac{2\pi L \alpha}{\lambda}\right) \, d\alpha \, d\beta \right)^{1/2} \\
   &= \frac{2\pi L \theta_S}{\lambda}  \sqrt{N_{s, \text{total}}}
\end{aligned}
\label{eq:2}
.\end{equation}

Since we have a basic priority scheme for the HZ around stars, we can set the baseline length to ensure the sensitive regions (the bright regions in Fig. \ref{fig3}) are located within the HZ. However, it's hard to know the exact position of planets; thus, we can only 
sample the HZ by adjusting baselines. Therefore, in the following part of this paper, we do not locate the planet at (0.89, 0.44) $\times \theta_p$, where the maximum S/N has been achieved (see Fig. \ref{fig3}). Instead, we fixed the planet at a position of (0.91, 0.41) $\times \theta_p$, which deviates from the ideal position in the simulations of the following sections.


\section{Effects of stellar activity on G dwarf stars}
\label{sec:3}
Various stellar activity parameters, including the position on the stellar surface and their activity intensity, can affect the depth of nulling interferometry. \textbf{Therefore, it influences the S/N when detecting Earth-like planets}. The S/N is given by Eq. \ref{eq:4}, \textbf{ where N\(_{act}\) represents the noise caused by stellar activity, as defined in Eq. \ref{eq:3}}. In this and the next section, we focus on the stellar activity around solar-like G dwarf stars and Proxima-like M dwarf stars, respectively, and study how the different flares influence the S/N of Earth-like planet. \textbf{Here, we adopt a fixed array to maximize the signal of Earth-like planets via interferometer via}
\begin{equation}
    \begin{split}
    N_{\text{act}} &= \iint_{S_{act}} (I_{act} - I_{s}) 4\sin^2\left(\frac{2\pi L \alpha}{\lambda}\right) \sin\left(\frac{4\pi \textbf{q}L \beta}{\lambda}\right) d\alpha d\beta \\
    &\approx \Delta N_{act}  \cdot 2\left(\frac{2\pi L \alpha_0}{\lambda}\right) ^3 \left(\frac{\textbf{q}\beta_0}{\alpha_0} \right)  \left[ 1  + \frac{1}{4}(\frac{\theta_0}{\alpha_0})^2 \right]
    \end{split}
\label{eq:3}
,\end{equation}

\begin{equation}
\begin{aligned}
\text{S/N} = \frac{N_{\text{planet}}}{\sqrt{N_{\text{poisson}}^2 + N_{\text{act}}^2}}
\label{eq:4}
\end{aligned}
.\end{equation}

\subsection{Position of active region}
\label{sec:3.1}
\begin{figure}[h]
    \centering
    \includegraphics[width=1\linewidth]{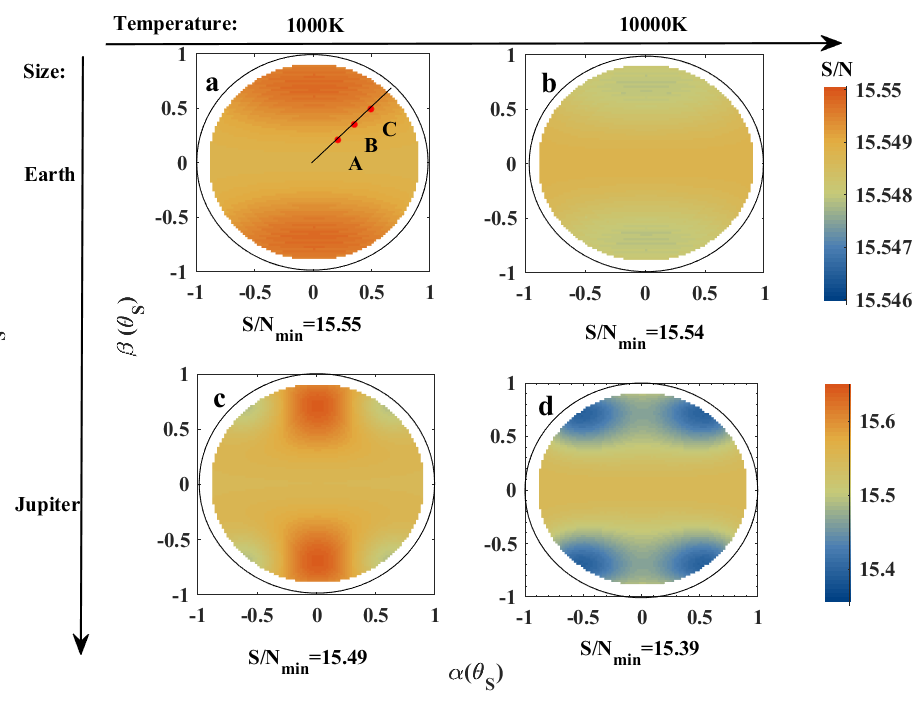}
    \caption{S/N values of planets when the stellar active region is located at different positions on G dwarf stars. Upper panels represent the Earth-size active region with temperatures of 1000 K and 10000 K for spots and flares, respectively. Bottom panels represent the Jupiter-size active region of spots and flares.  Points A, B, and C in panel (a) represent the position chosen to plot in Fig. \ref{fig6}.}
    \label{fig4}
\end{figure}
The position of the active region determines the optical path difference of the light received by different telescopes. According to Eq. \ref{eq:4}, the farther the activity region, $S_{act}$, centered at ($\alpha_0$, $\beta_0$) is from the center of the star, the greater the stellar leakage noise remains following the nulling interferometry. Here, ($I_{act}-I_{s}$) is the differential flux within the active region. Since the position of the stellar activity region does not extend beyond the stellar disk, the values of   \(\frac{2\pi L\alpha}{\lambda}\) and \(\frac{2\pi L\beta}{\lambda}\) are both less than \(3 \times 10^{-2}\) for G dwarf stars. Therefore, we can adopt $\sin x \approx x$, and the approximate expression is also shown in Eq. \ref{eq:4}, where $\Delta N_{act}$ is the variation in total photons within the active region. The center of the stellar activity region is located at \((\alpha_0, \beta_0)\), with an angular radius of \(\theta_0\). We note that Eq. \ref{eq:4} is only the noise induced directly by the activities, not including the variation in the Poisson noise due to the variations of stellar brightness. In our simulations, we considered the noise to be attributed to stellar activity, which includes changes in Poisson noise. The radiation areas of spots (1000 K) and flares (10,000 K) are set as two typical sizes for Earth or Jupiter, while the spatial positions of stellar activity are set at different locations on the stellar disk.

Based on the simulations in wide bands of 7-12 $\upmu$m, the relationship between planetary S/N and the position of the stellar activity region on the stellar surface was obtained, as shown in Fig. \ref{fig4}. We note that we consider the projection effect; namely,\textbf{ the projection effect refers to the change in the apparent size and shape of an active region on a star due to its rotation, even though the actual size remains the same.} The projected area of the active region becomes smaller when it is located further away from the center. As illustrated, the spots and flares at different locations can enhance and reduce the S/N while detecting habitable planets, respectively.\textbf{The reason is that spots and flares decrease and increase the Poisson noise, respectively. When active regions are small, the active noise, namely, the stellar leakage caused by stellar activity, is significantly smaller than the Poisson noise.} The Earth-sized spots, as shown in Fig.\ref{fig4} (a), were seen to reduce more Poisson noise when $\alpha_0 \sim \theta_s$ according to Eq. \ref{eq:2}; the S/N rises. Meanwhile, the Earth-sized flares lead to maximum Poisson noise or smallest S/N at $\alpha_0 \sim \theta_s$. We note that the effects of Earth-sized spots and flares on S/N are quite small and even negligible. 

When the region is Jupiter-sized, as shown in Fig.\ref{fig4} (c) and (d), the impact on the S/N becomes larger, especially when the position of the stellar activity region moves farther from the center of the stellar disk. The result is consistent with the positive correlation between the stellar noise and the position ($\alpha_0$, $\beta_0$) via both Eqs. \ref{eq:2} and \ref{eq:4}, \textbf{where q is the ratio of baseline and adopted as in Equation 1}.  In conclusion, the impact of both spots and flares on S/N is also not significant for G dwarf, with $\le$1\% reduction in S/N. 

Additionally, the influences are symmetric with respect to both the x-axis and y-axis. When the activity regions are near the edges of stars, the S/N values are highest when $\alpha_0=0$ for spots, and $\beta_0=0$ for flares. This is a combined effect considering both Poisson noise and the active noise. The S/N becomes smallest for Jupiter-sized spots and flares when $\alpha_0 \approx \beta_0$, which can be understood via the maximum stellar active noise, according to Eq. \ref{eq:4}.

\subsection{Stellar rotation}
\label{sec:3.2}
\begin{figure}[t]

    \centering
    \includegraphics[width=0.49\linewidth]{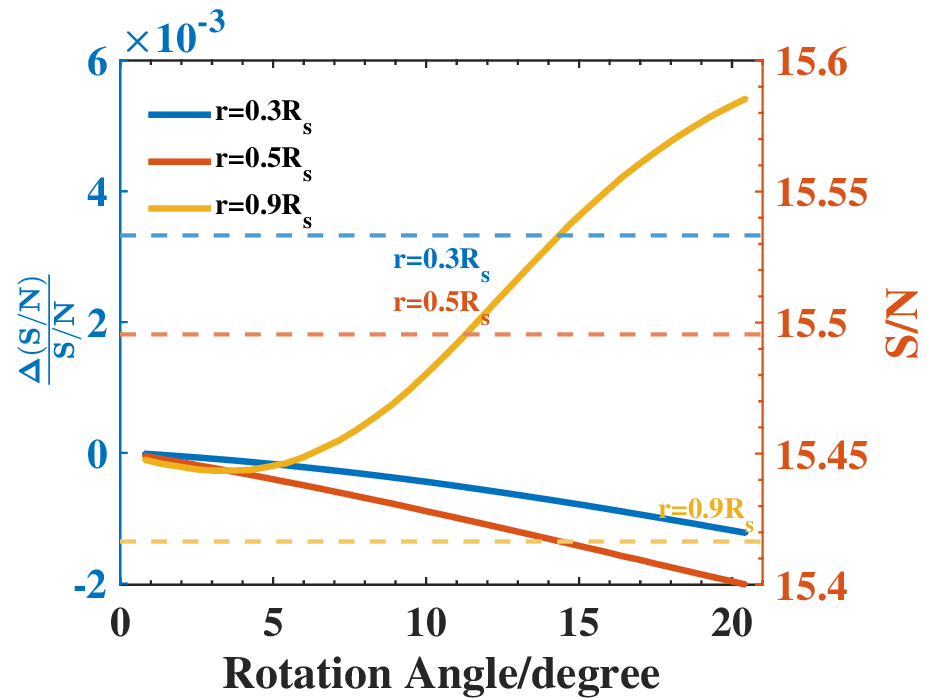}
    \includegraphics[width=0.49\linewidth]{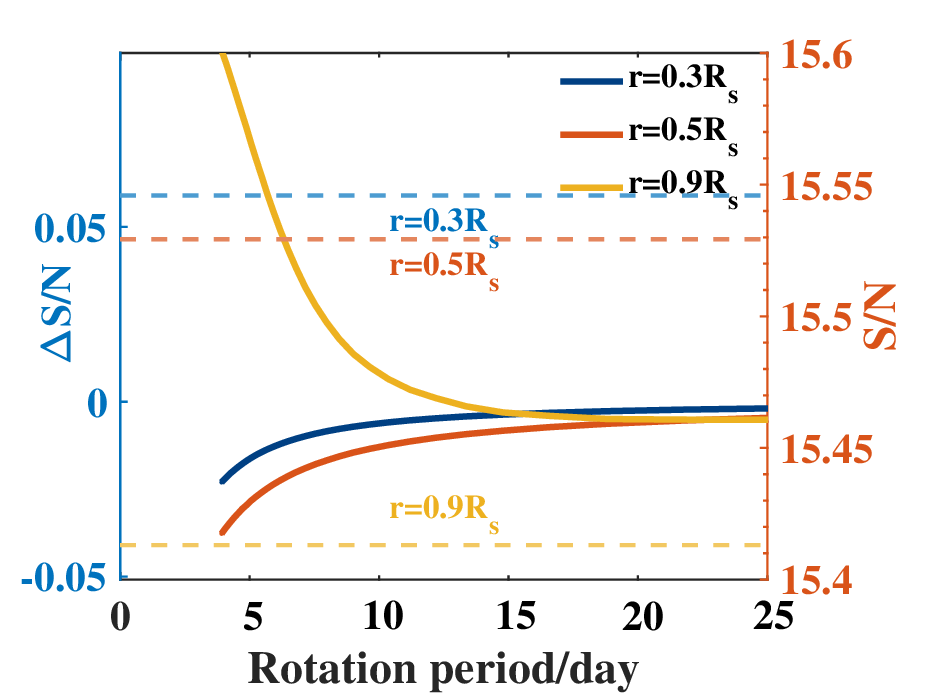}
\caption{The impact of stellar activity from a Jupiter-sized flare of 10,000 K, considering its position shift on the stellar disc during rotation. \textbf{ Left}: Effect of the Rotation Angle on S/N.  \textbf{ Right}: Variation in the S/N due to stellar rotation period by setting a total exposure time of 20000 sec. Here, $\Delta$S/N represents the difference between the S/N with and without stellar rotation, while S/N denotes the S/N in the presence of stellar activity without rotations. The dashed blue, red, and yellow lines indicate the S/N of the stellar activity region with different distances from the stellar center (i.e., 0.3, 0.5, and 0.9 $R_s$) by setting $\alpha_0=\beta_0$, respectively. The solid blue, red, and yellow lines represent\(\frac{\Delta \text{S/N}}{\text{S/N}}\)at the same distances.}
    \label{fig5}
\end{figure}

\textbf{ In Sect. \ref{sec:3.1}, we assume the star is non-rotating.} 
As the star rotates, the position of the activity region changes accordingly. When searching an Earth-like planet around the Sun at 10 pc with a S/N threshold of 10, a single measurement is set around 20000 sec, which is much shorter than the rotating period of the Sun (25.38 days sidereal according to \citet{25}); thus, the shift of stellar active region during one measurement is $3.3^\circ$, too small to affect the S/N significantly. For younger stars rotating faster, the active region can shift obviously and then lead to a considerable variation in the S/N. We tested different stellar rotating periods and saw the variation in the S/N based on comparisons with the non-rotating case, as shown in Fig. \ref{fig5}. We chose three positions as the initial locations of the active region, with a Jupiter-sized radius. The distances from the center were set as 0.3, 0.5, and 0.9\(\theta_{s}\) (setting $\alpha_0=\beta_0$). 

\textbf{As the stellar rotation period decreases, or the rotation angle becomes larger, the shift of the active region moves away from the center during the exposure time, and the suppression on the S/N of exoplanet detection becomes more significant. However, the suppression will turn to enhancement when the active region is very close to the edge. In the case of $r=0.9 R_s$, part of the active region can move to the backside and reduce the area in the observational view during the total exposure time; this results in an increasing S/N and can even keep a constant S/N after the whole active region moves to the backside.  Based on the effects of stellar rotation on S/N, precise measurements via nulling interferometry may also determine the position of active regions on rapidly rotating stars.}

\textbf{We note that for G dwarfs, the influence on S/N is not significant when considering single measurements (20000 sec), even if the rotating period is as short as 4 days. In reality, the nulling interferometry requires multiple measurements to cover the HZ around the star and determine the parameters of exoplanets. The continuous observational time on one target will be several tens of hours to a few days. For instance, when searching for an Earth-like planet around the Sun at 10 pc with a S/N threshold of 10, the total observational time should be $\sim$7 days \footnote{A group of five measurements with different baselines to find the planets in sensitive regions (see  Sect. \ref{sec:5}), and six groups (one divided by the efficiency), to cover the HZ. Thus, the total time cost is around $5\times6\times20000 sec\approx$ 7 day, without considering dead time.}, to cover the entire HZ around the star. Thus, the shift of active regions should be more obvious than a single measured effect due to stellar rotation. Typically, assuming the star’s rotation period is similar to the Sun  (5.38days sidereal \citep{25}), the active region will rotate across an angle of about $\sim 100^\circ$ in 7 days, that is the active region will shift by approximately 4/3 $R_{s}$ on the stellar disk. Therefore in most cases, the active regions can move to the backside. Therefore, the influence due to stellar rotation can be reduced when averaged over time. Based on the simulation in Fig. \ref{fig4}, when the activity region is Jupiter-sized with a temperature of 10000 K, the maximum reduction of the S/N is only around 1\%; specifically, the stellar leakage due to stellar rotation can only enhance the noise by 1\%. Compared with the S/N of the Earth-like planet around 15, i.e. the signal of the planet will be 15 times the noise, thus the stellar rotation may not influence the detection of Earth-like planets. However, when determining the location of the planet, the variation in the S/N will affect the accuracy (see Sect. \ref{sec:5}).}

\subsection{Size of active region}
\label{sec:3.2}

The size of the radiation area affected by stellar activity influences the observation of planetary signals. Theoretically, the larger the activity region is, the greater the noise generated. Assuming the position of the active region is constant, the variation in the stellar photons caused by stellar activity is affected by the area of the activity region, $S_{act}$, and the flux of the stellar activity, $I_{act}$, as indicated by Eq. \ref{eq:4}. The differential intensity within the stellar activity region is $\Delta N_{act}$, and the theoretical influences due to the size of the active region are shown in Eq. \ref{eq:4}. The second term in the bracket is a correction because of the finite size of active regions. It can be negligible when the size of the active region is small, namely, $\theta_{0}\ll \alpha_{0}$, and the noise is proportional to the area of the stellar activity region in the term of $\Delta N_{act}$. 

\begin{figure}[h]
    \centering
    \includegraphics[width=0.8\linewidth]{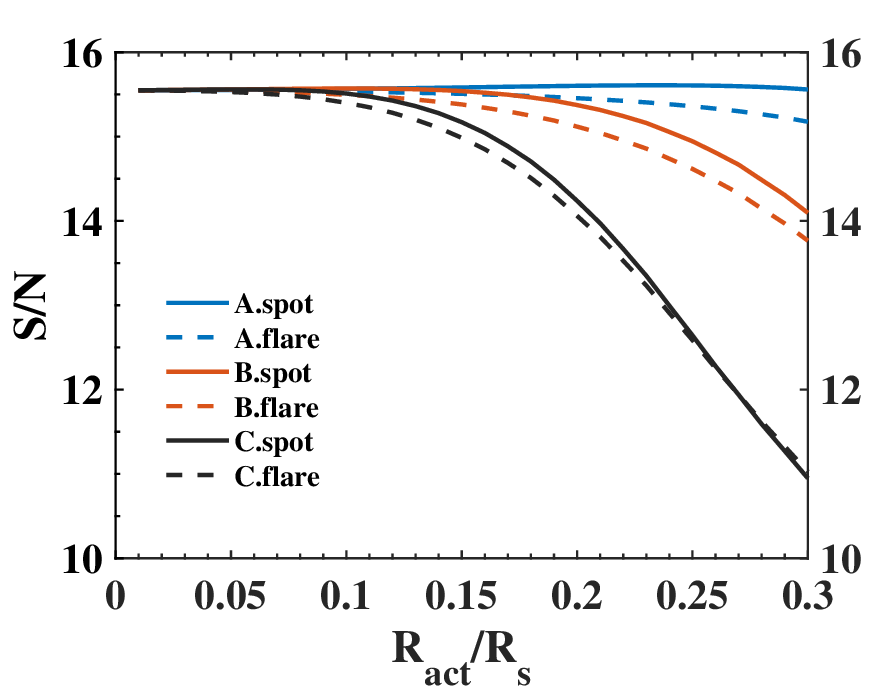}
    
    \caption{Relationship between the size of active regions and the S/N on G dwarf stars. The X-axis is the ratio of the radius of the stellar activity region to the stellar radius. Solid lines represent spots with temperatures of 1000 K. Dashed Lines represent flares of 10000 K. The symbols A, Bm and C represent the position of active regions, as shown in Fig. \ref{fig4}.}
    \label{fig6}
\end{figure}

To simulate the impact of stellar activity on planet detection, we chose  spots of 1000 K and flares of 10000 K. The sizes of the active regions are set from Earth-sized to three times the size of Jupiter. As shown in Fig. \ref{fig6}, the suppression of S/N by stellar activity becomes more obvious as the area of the activity region increases. For location A ($0.3\theta_{S}$ from the center), the reduction in the S/N is as low as <3\%, even if the active region covers 10\% of the stellar disk. For location B ($0.5\theta_{S}$ from the center), the S/N decreases to around 14, when the active region is three times Jupiter's radius. The S/N can even go as low as 11 for both the flare and spot, if the position moves to location C ($0.7\theta_{S}$ from the center), while the activity region is three times Jupiter's radius. 

Moreover, with the same area, the flares (10000 K) have larger influences on S/N than spot (1000 K), except for location C. For locations A and B, the noise due to stellar activities is smaller than Poisson noise. Because flares enhance the Poisson noise, while spots reduce the Poisson noise, the S/N due to flares are smaller than those due to spots. However, for location C, the active noise becomes comparable with Poisson noise, as well the projection effect also reduces the differences in Poisson noise due to flares and spots. Thus the influences of active noise due to spots or flares dominate, rather than the Poisson noise. Therefore the S/N becomes similar for spots or flares, according to the similar $\Delta N_{act}$ for flares and spots on Sun-like stars.

Even if the active region is near the edge of the stellar disk, it is clear that only an active region that is larger than twice the size of Jupiter may influence the S/N. For solar-like stars, which are usually not so active, most spots or flares are generated in areas smaller than Jupiter-size; thus, the influence on Earth-like planet detection is not crucial.

\subsection{Temperature of activity region}
\label{sec:3.3}
Besides the position and size of stellar activity regions, the intensity of stellar activity also influences the S/N of Earth-like planets. Assuming the blackbody radiation in the MIR band via the Planck formula, the temperature is a crucial parameter to determine the intensity of the stellar activity. For G dwarfs, the temperature of spots typically ranges from 1000 to 5000 K, while flares generally have temperatures between 9000 and 10000 K \citep{18,19,20,21}; therefore, we adopted a range from 500 K to 30000 K to cover the typical temperatures. We 
also adopted three typical sizes: 1, 3, and $10 R_{\oplus}$.

\begin{figure}[t]
    \centering
    \includegraphics[width=0.49\linewidth]{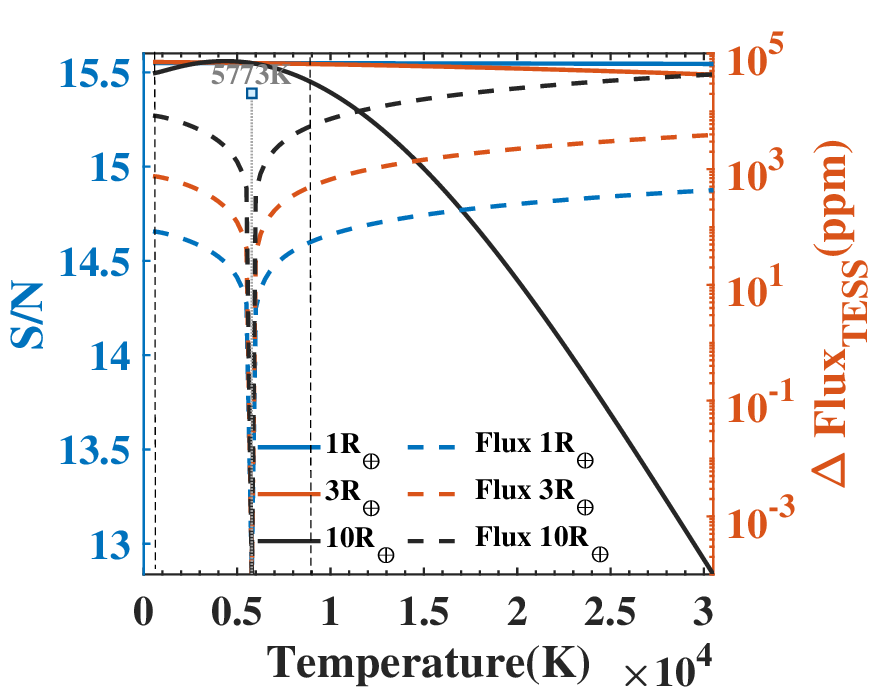}
    \includegraphics[width=0.49\linewidth]{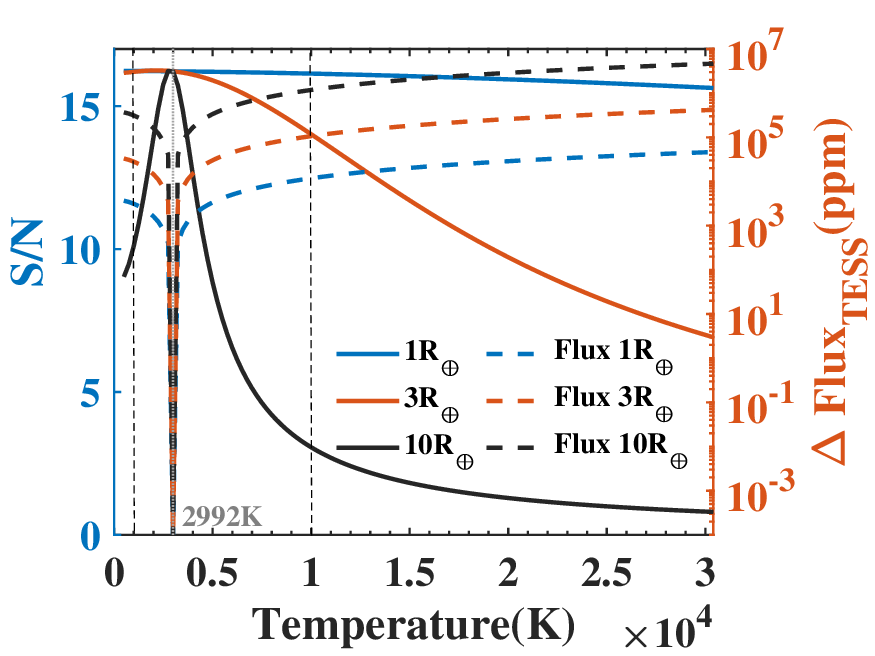}    
    \caption{Temperature of activity regions versus S/N and flux enhancement in the TESS observation band for active regions. The solid blue, red, and black lines represent the S/N of planet single with different active regions radii of 1, 3, and 10 $R_\oplus$, respectively. The dashed blue, red, and black lines indicate the flux enhancement due to stellar activity in the TESS observation band for active regions with the same radii. The left and right panels show the results on G and M dwarfs, respectively. The vertical lines represent the temperature of 1000 K (spots) and 10000 K (flares), as well as the stellar effective temperature.}

    \label{fig7}
\end{figure}

As shown in the left panel of Fig. \ref{fig7}, we can see that for solar-like stars, spots or flares with larger temperature deviations result in smaller S/N. When spot temperatures are as low as 1000 K and the active region is Jupiter-sized, the S/N is reduced by only $\sim 1\%$. For a Jupiter-sized stellar flare with a temperature of 10000 K, the S/N drops to \textbf{15.35}. If the temperature reaches 30000 K, the S/N drops to 15.5. For flares with 1 or 3 $R_{\oplus}$, even if the temperatures reach 30000 K, the influences on the S/N are negligible. Thus, most flares have very limited effects on the detection of Earth-like planets around G dwarfs.

To minimize the influence of stellar activity, especially the superflares, we can use the light curve of the star as reference data. Since the spots and flares can induce variations of the stellar brightness due to activity (as shown in the \textbf{black} dash lines of Fig. \ref{fig7} on the \textbf{left}), the Jupiter-sized flares, which reduce the S/N obviously from \textbf{15.5} to \textbf{12.86}, exhibit a normalized amplitude of around 4\% in the light curve in TESS band, which can be easily detected. Data collected during an active period can be removed when detecting Earth-like planets via interferometry. \textbf{We note that the S/N considerations presented in this paper are valid only for a fixed array configuration with an optimized baseline without rotation. With an array rotation, the planet signal and the signals from active regions will be modulated at different frequencies. In particular, when  stellar activity continues on a long timescale, the stellar activity noise can be removed via array rotation based on synchronous demodulation, thanks to the lower frequency due to activities.}

\section{Effects of stellar activity on M dwarf stars}
\label{sec:4}
In Sect. \ref{sec:3}, we consider the influence of stellar activity on G dwarfs. However, M dwarfs dominate the nearby stellar neighborhoods, especially in the solar one \citep{248}. Dozens of planets in the HZ have been detected around M dwarfs and the CARMENES exoplanet survey of M dwarfs indicates a higher occurrence rate of Earth-like planets around M dwarfs \citep{249}. In this section, we mainly focus on the influence of stellar activity on detecting Earth-like planets around M dwarfs, such as Proxima Centauri at \textbf{10 pc}.

\begin{figure}[t]
    \centering
    \includegraphics[width=1\linewidth]{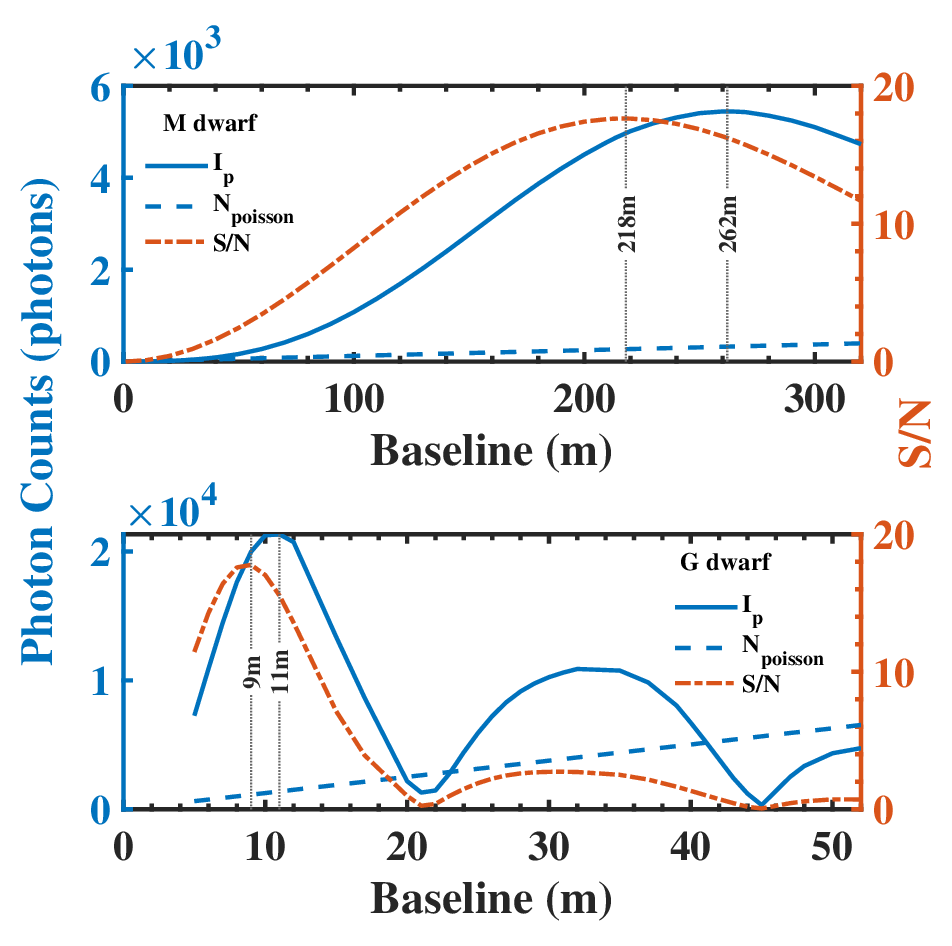}
   \caption{\textbf{Variation in the planetary signal (solid blue line), stellar Poisson noise (dashed blue line), and the S/N (dashed red line) as the baseline length changes. The gray dotted lines indicate the baseline lengths for M-dwarfs (top) and G-type (bottom) stars at which the signals for detecting habitable planets or the S/N are at maximum. Note: the total exposure time for a M dwarf is 5000 sec, compared to 20000 sec for a G dwarf.}}
    \label{fig8}
\end{figure}

To make a comparison with the results of G dwarfs, here we strengthen the significant differences between M and G dwarfs. First, as shown in Fig. \ref{fig2}, M dwarfs generally have lower luminosity, especially in the middle infrared band; thus, the contrast between star and planet would be lower. As well, the Poisson noise of M dwarfs is significantly smaller than G dwarfs. The stellar Poisson noise after nulling is determined by the factor of $L*\theta_S$. Since the HZ around M dwarfs are much closer than G dwarf stars, the baseline is much longer. Fig. \ref{fig8} shows the Poisson noise of a star (dashed blue line), the signal of the planet (solid blue line) in \textbf{differential} output, and the S/N of the planet (red dash-dot line), without activities. \textbf{Taking a similar approach to a  G-type star, we adjust the orientation of planets and the baseline to maximum the signal of planets, then we fixed the array to investigate the influence of activities.}  The maximum planet signals are achieved when the baselines are set as 11 meters and 262 meters for G and M dwarfs. Since the stellar Poisson noise \textbf{increases} with baselines, the maximum S/N can be achieved on shorter baselines. When substituting the typical values in Table \ref{table1}, the factor of $L*\theta_s$ for an M dwarf is 3.5 times larger than for G dwarfs. In combination with the much less stellar photons known for M dwarfs, the Poisson noise of M dwarfs (according to Eq. \ref{eq:2}) is only $\sim 47\%$ of G dwarfs if the total exposure time is the same. Therefore, detecting Earth-like planets around M dwarfs seems much easier, if we do not consider the stellar activities. To make sure the maximum S/N is comparable for M and G dwarfs, we adjusted the total exposure time to 5000 sec for M stars. 

Secondly, M dwarfs are usually more active \citep{a2}, especially the fully convective M-star \citep{a3}. The coverage fraction of stellar active regions is 10\% and 30-50\% for G and M dwarfs, respectively \citep{a4}. For Jupiter-sized active regions, the coverage on the stellar disk is around 10\% and 50\% for solar-like stars and Proxima-like stars. Therefore, we also chose the size of active regions from Earth-sized to Jupiter-sized. Additionally, the temperature of flares on M dwarfs could be higher than 30,000 K \citep{a6}, much hotter than those on G dwarfs; thus, there would be some superflares on M dwarfs, which was observed on Proxima \citep{a7}.

\subsection{Position of activity region}
\label{sec:4.1}

\begin{figure}[h]
    \centering
    \includegraphics[width=1\linewidth]{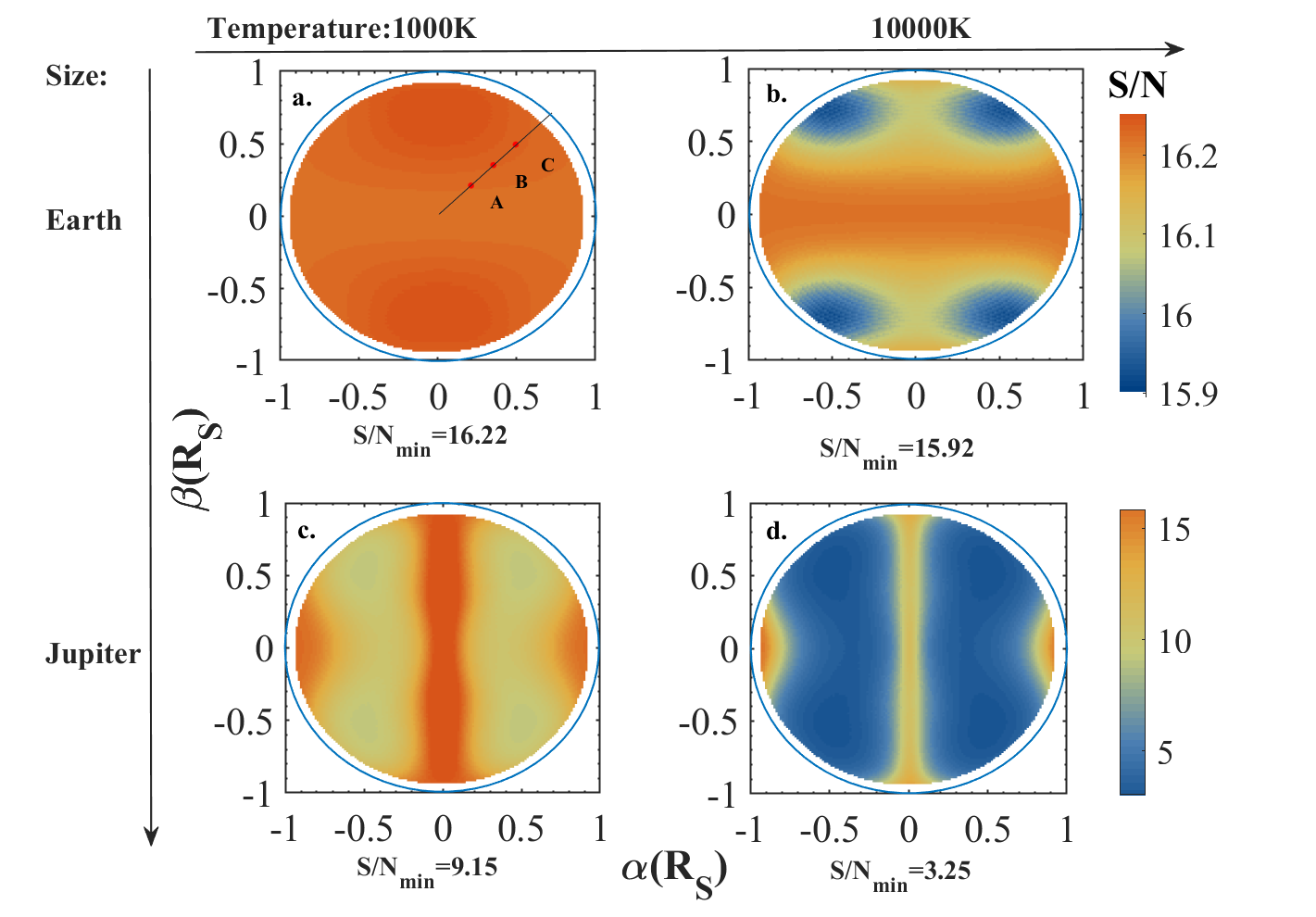}
    \caption{Effect of position of stellar active region on M dwarf stars. Panels (a)-(d) represent the S/N when the active region is in different positions. The upper panels represent the Earth-size active region with temperatures of 1000 K and 10000 K, for spot and flare, respectively. The bottom panels represent the Jupiter-size active region of spots and flares.
    Points A, B, and C in panel (a) represent the position chosen to plot in Fig. \ref{fig10}.}
    \label{fig9}
\end{figure}

First, we studied the influence due to the position of stellar active regions via simulations, as shown in Fig. \ref{fig9}. Compared to G dwarf stars, the impact of stellar activity of the same size and temperature is much stronger for M dwarfs. This is mainly because: 1) the size of M dwarfs is 0.15$R_{\odot}$ and if the size of active region keeps the same with G dwarfs, the larger asymmetry of flux on the stellar disk is caused by flares or spots; and 2) the interferometric configuration for detecting M dwarfs requires longer baselines and the active regions near the edge of stellar disk have a larger contribution to the noise, according to Eqs. \ref{eq:2} and \ref{eq:4}. 

The variation in S/N values with different positions is similar to Fig. \ref{fig4}, when the active region of spots is small (Earth-size), because of the Poisson noise. However, in panel (b), flares that are related to Earth-sized areas can lead to minimum S/N when $\alpha_0\approx\beta_0$, similar to the pattern of a panel (d) in Fig. \ref{fig4}. This is because the active noise becomes larger than the Poisson noise, due to the larger normalized intensity of activity related to the M dwarf. For Earth-sized active regions, including both spots and flares, the decrease in the S/N is insignificant and does not influence the detection of Earth-like planets. 

For Jupiter-sized spots (panel (c) in Fig. \ref{fig9}),  the smallest S/N is around 11 and it is not serious for the detection of Earth-like planets. The variation in the S/N seems more complex. The S/N decreases as the stellar activity region moves farther from the center, except in the direction of two axes. In the direction of $\alpha$, since the active noise increases with $\alpha_0$ according to Eq. \ref{eq:4}, meanwhile the Poisson noise decreases according to Eq. \ref{eq:2}. The projection effect is the same for both noises, thus, the S/N remains nearly constant. In the direction of $\beta$, the spot will not change the Poisson noise with different $\beta_0$. Both Poisson noise and active noise are mainly reduced because of the projection effect, especially when part of the spots move to the backside. Therefore, the S/N increases when $\beta_0$ becomes closer to $\theta_s$. A similar pattern is also shown in panel (d) in Fig. \ref{fig9} for flares with the same reasons. Additionally, the S/N can be largely reduced to \textbf{3.25} for such a superflare. Thus the measurements on M dwarfs via nulling interferometry should be careful to exclude data obtained during super flares.

\subsection{Size of activity region}
\label{sec:4.2}
\begin{figure}[h]
    \centering
    \includegraphics[width=1\linewidth]{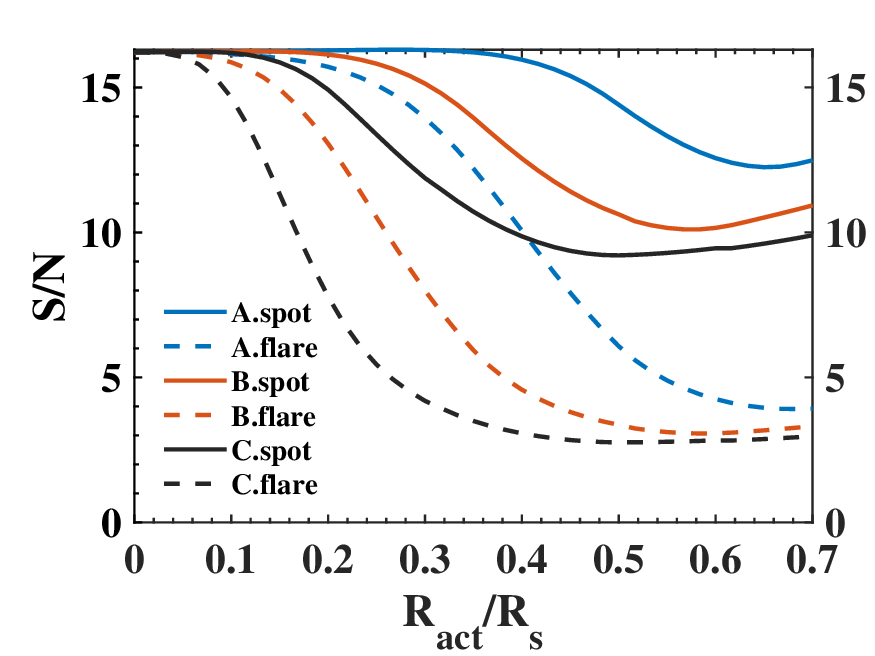}
      \caption{Relationship between the area of stellar activity regions and observational S/N on M dwarf stars. The X-axis is the ratio of the radius of the stellar activity region to the stellar radius. The solid lines represent spots with a temperature of 1,000 K, while the dashed lines represent flares of 10,000 K. Symbols A, B, and C represent the position of active regions, as shown in Fig. \ref{fig9}.}
    \label{fig10}
\end{figure}

To estimate influences on the S/N due to different sizes, we choose three positions A, B, and C, in the top-left panel in Fig. \ref{fig9}. The typical temperature of spots and flares are set as 1000 K and 10,000 K, respectively. As shown in Fig. \ref{fig10}, the suppression of S/N by stellar activity becomes more pronounced as the radiation area of the stellar activity region increases. For location A (i.e 0.3 times the stellar radius from the center), the impact of flares is obvious when the active regions are larger than $0.44R_{s}$ ($\sim 6.6 R_{\oplus}$), reducing the S/N from \textbf{16 to 8}. For locations B and C (i.e.,  0.5 and 0.7 times the stellar radius from the center), the impact of flares decreases the S/N to below 10 when the radius is larger than $0.3 R_{s}$ and $0.2 R_{s}$, respectively. For some superflares of a Jupiter size, the detection of planets seems difficult due to the low S/N ($\sim$3.9). The influence of spots is much smaller than flares because the temperature difference between spots and the M dwarf is much smaller than flares. Even Jupiter-sized spots can sustain a S/N$>$10. 

Additionally, the S/N does not always decrease when the size becomes larger. The S/N due to flares shows a slight increase when the size becomes larger. Since the stellar disk is only 1.5 times Jupiter's Radius, in the case of the active region centered at C, if the size is larger than 0.5 $R_{s}$, the active region can cover part of regions with $\beta<0$. This will reduce the asymmetry of the stellar disk. For example, when combining the projection effect, part of the activity noise can be eliminated by the odd-parity of $\beta$ in Eq. \ref{eq:4}. The S/N due to the spot also increases when the size is close to Jupiter's size. Besides the reducing asymmetry, the reduced Poisson noise also enhances S/N, thus the increment of S/N with large sizes is slightly more obvious than flares.

\subsection{Temperature of activity region}
\label{sec:4.3}
 As shown in the right panel of Fig. \ref{fig7}, we can see that the lower the temperature of the spots, the greater the impact on the S/N. When the active area is as large as Earth, the impact of stellar activity on the S/N is small. Even for flares with 30000 K, the influence of Earth-sized active areas can reduce the S/N by around 5\%. Thus, the active noise due to a small active area (Earth-sized) is negligible. When the flare radiation area is $3 R_{\oplus}$ with temperature $>20000$ K, or Jupiter-sized with temperature $>5200$ K, the S/N can be reduced to 10. The superflares on M dwarfs can influence the detection of Earth-like planets. In the extreme case of Jupiter-size flares with 30000 K, the S/N is as low as $\sim 1$. We note superflares can be easily detected via photometric data. For the extreme case, the enhancement of flux in the TESS band would be $> 400\%$. Thus, in combination with photometric data, this allows us  to rule out the data as outliers during the superflare events detected on M dwarfs. With a criterion of $<40\%$ variation in  flares on M dwarfs in the TESS band, the influence of activities is limited and S/N is $> 10$ to guarantee the detection of Earth-like planets.

 As a conclusion to this section (albeit without considering noise from stellar activity), the detection of exoplanets around M dwarfs achieves a higher S/N compared to G dwarf stars. However, the S/N can be easily influenced by stellar activities. Considering that M dwarfs are usually more active and the relative radiation intensity of activity on G dwarf stars is usually lower than that on M dwarfs, the impact of stellar activity is more significant in observations around M dwarfs via interferometry. We can refer to the photometry data to exclude some superflares in some active periods of stars. Also, the selection of inactive M stars is important for searching the Earth-like planets around nearby M dwarfs. 


\section{Influences on planetary position measurements}
\label{sec:5}

\begin{figure}[h]
    \centering
    \includegraphics[width=1\linewidth]{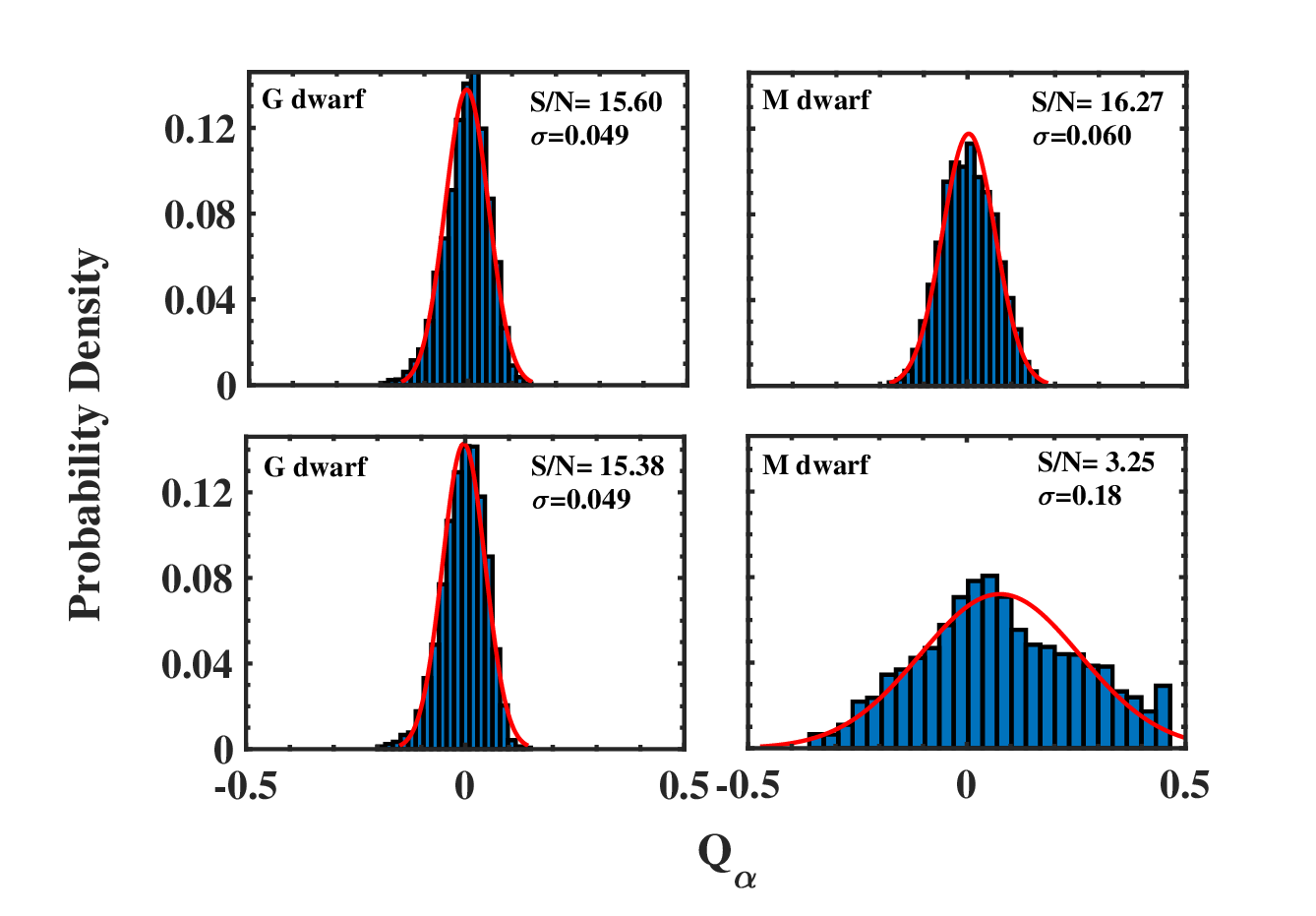}
   {\caption{Distribution of the best fitted relative accuracy of planet position $Q_\alpha$ for G dwarfs (left) and M dwarfs (right). The probability density indicates a symmetric Gaussian distribution. The red line represents a Gaussian fit of the relationship between $Q_\alpha$ and the probability density. The S/N values adopted in four panels corresponds to the maximum and minimum S/N given in Figs. \ref{fig4} and \ref{fig9} for G and M stars, respectively.}
}
    \label{fig11}
\end{figure}

\begin{figure}[h]
    \centering
    \includegraphics[width=1\linewidth]{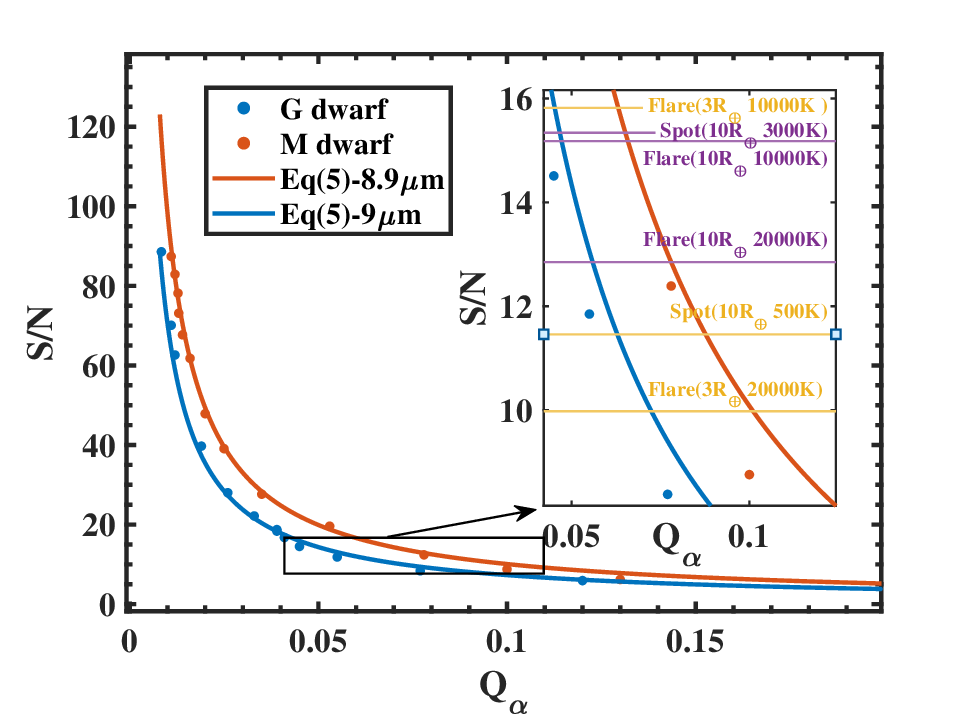}
   \caption{Relationship between the impact of stellar activity on the S/N for detecting exoplanets around G dwarf (blue points) and M dwarf stars (red points) and the positional measurement error $Q_\alpha$ of the planet. The solid line indicates the theoretical value from Eq.\ref{eq:6} at 8.9 $\upmu$m (red) and 9 $\upmu$m (blue). The horizontal solid lines correspond to the S/N of detecting exoplanets around G dwarf (purple) and M dwarf (yellow) for stellar activity with a specific radiation region radius and intensity.}
    \label{fig12}
\end{figure}

When detecting habitable planets, it is not only crucial to identify the presence of a planet, but also to confirm that the planet is indeed located within the HZ. Assuming a criterion of 1-$\sigma$ confidence level to judge whether a given  planet is located in the HZ, an approximate relationship between the relative position precision \( Q \) (i.e., the $1-\sigma$ uncertainty divided by the real value) of the planet and the probability \( P \) is given by Eq. \ref{eq:5}. We note that \( P \) represents the probability of planets located in HZ randomly, considered within the HZ according to the fitted positions, including the 1-$\sigma$ uncertainty of their positions. The $F$ value is determined by the HZ around stars, that is, $F=(a_{HZ,in}+a_{HZ,out})/(a_{HZ,in}-a_{HZ,out})$. This gives us 3.34 for solar-like stars and 2.48 for M dwarfs such as Proxima. To ensure a probability  of at least $80\%$ that a planet may be accurately considered to be within the HZ, the required relative precision of the planet's position is 6\% for G dwarfs and 8\% for M dwarfs, as per
    
\begin{equation}
\\ \\ Q=(1-P*F)
\label{eq:5}
.\end{equation}

To obtain the relations between the S/N and the 1-$\sigma$ uncertainty of planet's position, $Q$, we derived Eq. \ref{eq:6} for fixed wavelength according to Eqs. \ref{eq:1} and \ref{eq:2}. We can see that the precision in two different directions is not the same and depends on the position of the planet $(\alpha,\beta)$. To check the accuracy of the theoretical relationship, we also fit the position of planets via simulations in a wide band of 7-12 $\upmu$m, as per

 \begin{equation}
\text{S/N} \approx \frac{\lambda}{4\pi L}\times \frac{1}{\alpha \cot\left(\frac{2\pi L \alpha}{\lambda}\right) Q_{\alpha} + \beta \cot\left(\frac{4\pi \textbf{q}L \beta}{\lambda}\right) Q_{\beta}}
\label{eq:6}
.\end{equation}

To simplify the simulation, we only considered the position in the direction of $\alpha$. We also fixed a planet at \((0.91, 0.41) \times \theta_p\) for both G and M dwarfs, where the S/N of the planet is close to the maximum value. The S/N is set as a free parameter, which impacts the precision \( Q _{\alpha}\) of the planet's position. We selected five baseline lengths, namely: 9, 10, 11, 12, and 13 meters for the G dwarfs and 222, 242, 262, 282, and 303 meters for the M dwarfs. By sampling the outputs and adding a Gaussian noise according to the S/N 5000 times, we fit the planetary position in the direction of $\alpha$ and compare it with the fixed planetary position. Thus, we were able to obtain the relative deviation of fitted planet position, $Q_{\alpha}$. As shown in Fig. \ref{fig11}, the distribution of fitted $Q_\alpha$ for G and M dwarfs are all Gaussian, except the M stars with S/N as low as \textbf{3.25}.

The relationship between S/N and $Q_{\alpha}$ is shown in Fig. \ref{fig12}. It is consistent with the trend in Eq. \ref{eq:6} for both stars. When the relative error in the planet's position measurement $Q_{\alpha}$ is 6\% and 8\% (corresponding to 80\% probability that the detected planet is within the HZ), the required S/N for detecting exoplanets around G dwarf and M dwarfs is 12 and 14, respectively. Based on the simulation results from Sect. \ref{sec:3} and Sect. \ref{sec:4}, if we were to consider the impact of stellar activity, the S/N values of 12 and 14 correspond to a Jupiter-sized flare (20000 K) on a G dwarf and a Jupiter-sized spot (1000 K) on a G dwarf star, respectively. Thus, to guarantee the probability $P \geq 80\%$, we can constrain the stellar activities according to the simulated results. For the detection threshold of S/N$=$10, the $Q_{\alpha}$ of G and M dwarfs are  7\% and 10\%, respectively, corresponding to the probability of 76.7\% and 75.2\%.

 

The differences in the relationship between G dwarf and M dwarf in Fig. \ref{fig12} are caused by the different stellar spectrum, that is, the contribution of flux at a shorter wavelength on G dwarf is larger, thus, the influence with the same S/N leads to a smaller $Q_{\alpha}$ according to Eq. \ref{eq:6}. A similar trend of $Q_\beta$ can be obtained and Eq. \ref{eq:6} is also available for M dwarfs. We note when the S/N is lower, Eq. \ref{eq:6} is not accurate because the higher order of $Q$ will influence the accuracy of Eq. \ref{eq:6}. 

Compared with the detection of Earth-like planets, active noise is more likely to reduce the probability of determining if the planet is in the HZ. For G dwarfs, the spots or flares usually lead to an uncertainty smaller than 0.05. Even for superflares (10$R_\oplus$, 20000 K), the uncertainty is around 0.06, while the probability $P\approx81\%$. Therefore the activity noise is still tolerated. However, typical activities on M dwarfs (i.e.  the flare or spot) will lead to a hight result, $Q_\alpha>0.1,$ and reduce the probability to $P<75\%$. Therefore, to achieve more accurate measurements of a planet’s position around M dwarfs, it is necessary to combine photometric observations and exclude  data obtained during periods of intense stellar activity. 


\section{Discussion}
\label{sec:6}

\begin{figure}[h]
    \centering
    \includegraphics[width=1\linewidth]{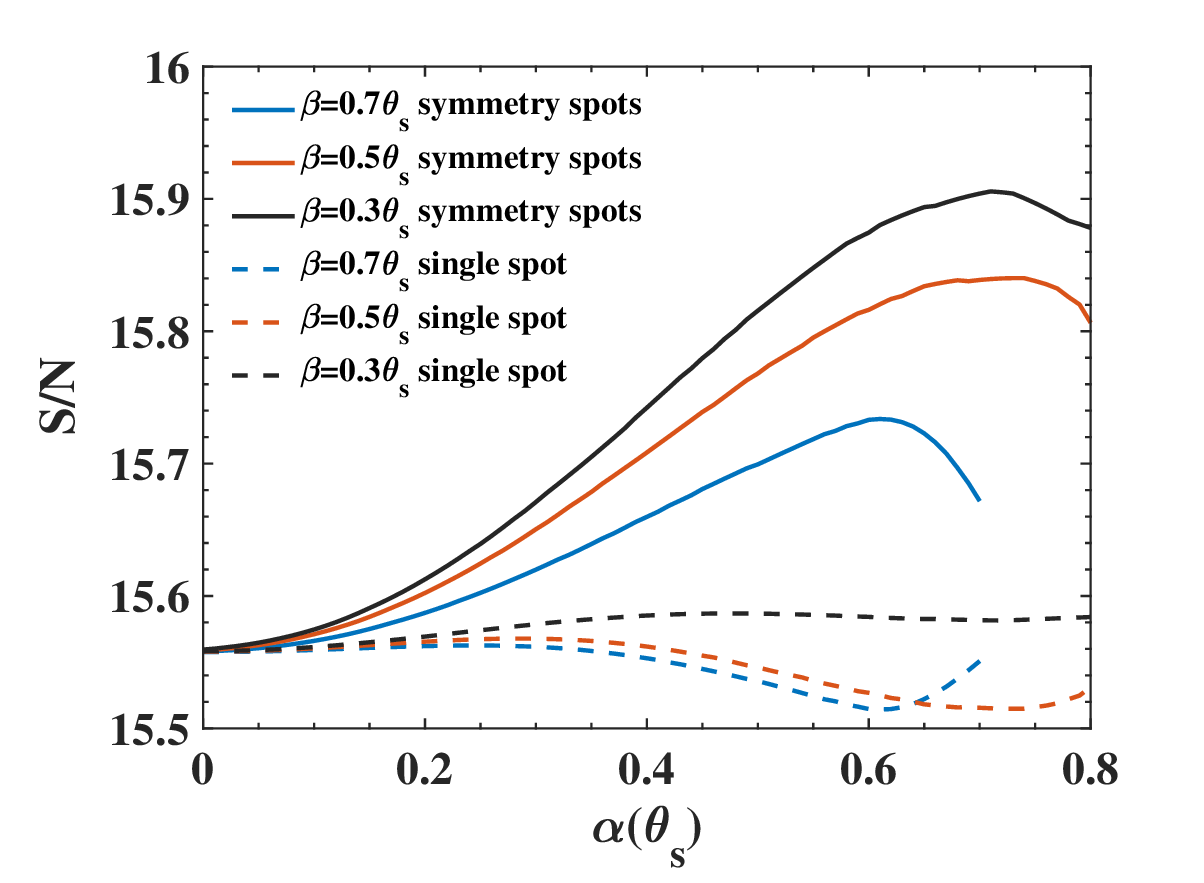}
    \caption{Impact of a single spot (dashed lines) and two symmetrical spots (solid lines) on the S/N. Here, $\alpha$ and $\beta$ represent the positions of spots on solar-like star, and $\theta_s$ is the stellar angular radius. The temperature of the spots is fixed as 1000 K and the radius is Jupiter-sized.}
    \label{fig13} 
\end{figure}

Since we are focused here on the stellar activity of G dwarf and M dwarf stars, we only considered a single active region. However, there might be a group of active regions that exist simultaneously on the stellar disk. As observed on the Sun, sunspots usually exhibit a roughly symmetrical spatial and temporal distribution across the northern and southern hemispheres, a phenomenon known as the "butterfly diagram" \citep{421}. Considering the symmetrical structure of active regions, we simulated the influence of two symmetric spots (1000 K) with Jupiter-size on the disk of G dwarfs, as shown in Fig. \ref{fig13}. Different from the single spots, the S/N increases when the symmetry spots move away from the center. This is because the active noise is eliminated by symmetry and the Poisson noise dominates, which decreases when spots move further from the center in the direction of $\alpha$. However after the spot moves to near the edge of the stellar disk ($\sim$0.6-0.7$\theta_{S}$), the projection effect reduces the influence of spots and then the Poisson noise becomes higher to make the S/N decrease again. Gratifyingly, the stellar activity is not always awful but may benefit detecting Earth-like planets if the spots are symmetric.

Moreover, there are other noises which would couple with the stellar activities. For instance, the pointing error as well as the \textbf{instability noise} will make the stellar disk not always in the center. The influence of stellar activity also depends on the pointing accuracy of the nulling interferometry, according to Eq. \ref{eq:4}, where the active region $S_{act}$ is no longer centered at (0,0). Oliver et al.\citep{15} have studied the impact of instability on the direct detection of exoplanets with nulling interferometers. Besides, some other astrophysics noises are also crucial for detecting Earth-like planets. \textbf{Exozodiacal light, which can interfere with terrestrial planet detection \citep{8}, has been measured around nearby stars using nulling interferometry by Ertel et al. {\citep{R5}} and shown to be sufficiently low. According to the study from \citep{R4}, its impact can be tolerated on the X array.}

In this paper, both spots and flares on
the stellar disk is considered. The stellar activity can also lead to coronal mass ejection (CME). Although CME is faint in MIR, but the speed can be up to 1000 km/s \citep{388}, and last around 10 hours \citep{389}. Thus the CME can go as far as ten times the stellar radius. As shown in Eq. \ref{eq:4}, the asymmetric stellar activity signal further away from the center leads to larger noise, the influence of CME might be not negligible,\textbf{and will require further studies}.

We use typical G and M dwarf stars with a single Earth-like planet, to study the uncertainty of the planet's location due to stellar activities. For multiple planetary systems, the retrieval of planetary parameters at different locations will have different accuracy and S/N \citep{8}; therefore, the influence of stellar activity on their position determination is more complex and requires a   case-by-case study; in particular for the terrestrial planets around M dwarfs, which are more compact and even have several rocky planets in the HZ. Furthermore, the strong stellar flares of M dwarfs may also induce the variation in the planet's atmosphere \citep{392-1,392-2}. Consequently, the thermal emission of the planet through the atmosphere will also vary, which leads to additional uncertainty of planetary signals, as well as the locations. Thus it would be better to detect Earth-like planets around G dwarfs, which have less of an effect on planets, rather than M dwarfs.

\begin{figure}[h]
     \centering
    \includegraphics[width=1\linewidth]
    {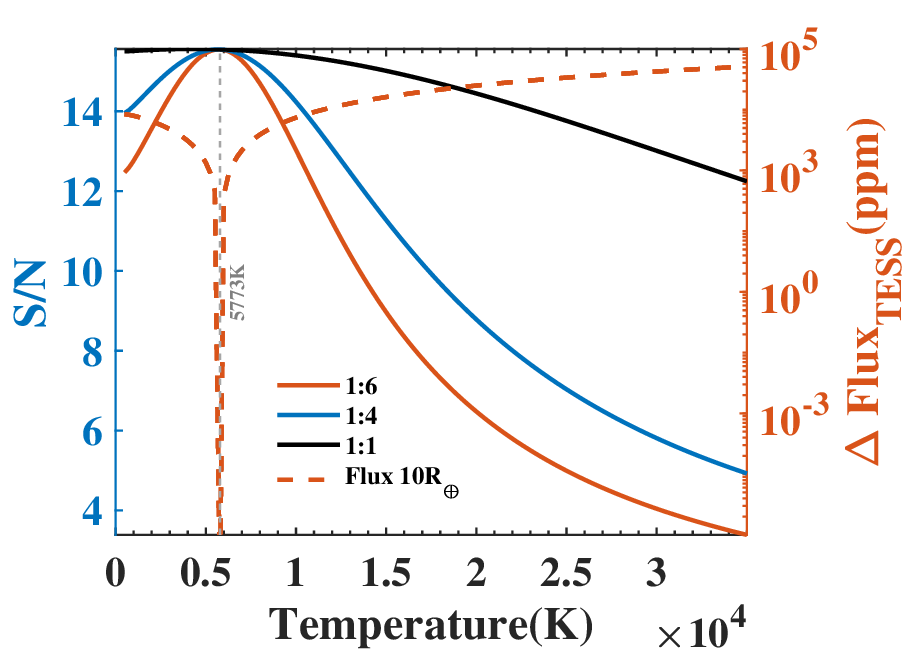}
    \caption{Impact of stellar activity on the S/N for detecting habitable planets in G-type star systems.Line: The S/N for detecting habitable planets with the X array when the baseline length ratio is 1 (black), 4 (blue), and 6 (red). Red dashed line:  Relative flux of stellar activity. The active regions have a radius of \(10 R_\oplus\), and are fixed at ($0.5 R_S$,$0.5 R_S$).}
    \label{fig14} 
    \end{figure}

\textbf{Furthermore, we considered the case with an aspect ratio of 4 or 6 \textbf{ to mitigate the instability noise} (as proposed by Lay 2006). Compared with the 1:1 configuration, when the stellar activity has the same intensity, size, and position, the S/N becomes lower as shown in Fig.\ref{fig14}. It is consistent with the results in Eq. \ref{eq:3}; namely, the noise caused by stellar activity increases by a factor of \( q \), when the baselines are extended to q times in one direction, as shown in Fig. \ref{fig3}.}


\section{Summary and conclusion}
\label{sec:7}
In this paper, we investigate the influence of stellar activity on the direct detection of exoplanets in a wide MIR band (7-12 $\upmu$m), using an X array configuration nulling interferometer (see Table \ref{table1}). Stellar activity, which includes dark spots and bright flares on the stellar surface, introduces additional noise in interferometric measurements, including variations in the Poisson noise and the activity noise that complicate the detection of exoplanets. In Sect. \ref{sec:2}, we introduce the spectral models of the planet, stars, and active regions used in this paper. We adopted an optimized 1:1 X-configuration \textbf{with fixed baselines and orientations} to investigate the influence of S/N due to spots and flares on two typical planetary systems; namely, Earth-like planets around Sun-like G-dwarf and Proxima-like M-dwarf at 10 pc (see Table \ref{table1}).

In Sect. \ref{sec:3}, we derive Eq. \ref{eq:4}, which demonstrates the influences due to stellar activities, and find that the location and the intensity of these active regions are critical factors affecting the S/N values of Earth-like planets. Based on the simulated results for typical solar-like G-dwarfs (see Table \ref{table1}), we investigated the influence of the position, size, and temperature of active regions. We see how these different activities impact the S/N of an Earth-like planet around a solar-like star. The Poisson noise and active noise were added together, along with  the projection effect; specifically, the projected area becomes smaller when the active regions move from the center to the edge. Most activities have negligible impact on S/N, while only Jupiter-sized flares with very high temperatures $>20000$ K can suppress the S/N in a limited way. Additionally, spots on G dwarfs can enhance the S/N slightly by reducing the Poisson noise, while flares will reduce the S/N by raising both Poisson and active noise. Considering the star rotations, the influence of stellar activity is also negligible, because when the active regions are near the edge of the stellar disk, the noise contribution may move to the backside due to rotation.

In Sect. \ref{sec:4}, we focus on M dwarfs similar to Proxima (see Table \ref{table1}). The study finds that M-dwarfs generally offer higher S/N than G dwarfs under ideal conditions, primarily due to their lower star-planet contrast. We can also achieved a S/N of around \textbf{16} by reducing the total exposure time from 20000 sec to 5000 sec. The S/N reduction due to stellar activity is more pronounced around M-dwarfs, which are prone to larger relative variation and more asymmetric stellar disks due to activities. Especially for some superflares, that is, Jupiter-sized flares with temperatures of >10000 K can reduce the S/N to <\textbf{3.25}. Since the flares on M stars are usually more frequent than G stars, we should deal with the interferometric measurements with caution. One possible way is to combine the photometric data to exclude the influence of superflares. For instance, flares on M stars with relative amplitude $<40\%$ in the TESS band, which can be easily detected, can guarantee a S/N>10 for detecting Earth-like planets. 

In both the cases of G and M dwarfs,  stellar activities have much less of an influence on detecting Earth-like planets, as compared to both the transit and RV. However,  stellar activities can influence the accuracy of determining the orbital parameters of planets in HZ, even at S/N > 10. In Sect. \ref{sec:5}, we explain how we obtained Eq. \ref{eq:5} to represent the probability, $P$, to retrieve a planet inside the HZ, considering an uncertainty of a position, $Q$. Then, we derive Eq. \ref{eq:6} to approximately represent the relationship between the S/N and the relative uncertainty of the planet's position. Our simulated results fit  Eq. \ref{eq:6} well, as shown in Fig. \ref{fig12}, when $Q_\alpha<0.1$. With a threshold of S/N>10, the probability of correctly identifying an Earth-like planet around G and M dwarfs is 76.7\% and 75.2\%, respectively.

In conclusion, the results in this paper quantitatively demonstrate the influence of stellar activities  and indicate the less interference of detection of Earth-like planets around different stars, due to the \textbf{X-array} nulling interferometer. However, we prefer detecting planets around solar-like stars or inactive M stars with higher priorities. This is because of the uncertainties of the unknown planetary atmosphere, which can be totally eroded by the strong activity seen for  active M dwarfs. The influence of stellar activities on the S/N can be also utilized to characterize the position, size, and temperature of the spot or flare on the stellar disk, when we are able to obtain higher S/N. In the future, we can also refer to the reference data \textbf{to calibrate the output} and reduce different types of noise, including systematic noise and background noise, as well as stellar noise, to achieve a more clear signals from  Earth-like planets.

\begin{acknowledgements}
This work is supported by the National Key R\&D Program of China (2019YFA0706601) and the National Natural Science Foundation of China (grant Nos. 11973028, 11933001, 1803012, and 12150009). We also acknowledge the science research grants from the China Manned Space Project with Nos. CMS-CSST-2021-B12, as well as the Civil Aerospace Technology Research Project (D010102).
\end{acknowledgements}

%
%

\bibliography{aa52725-24}

\begin{thebibliography}{49}
\expandafter\ifx\csname natexlab\endcsname\relax\def\natexlab#1{#1}\fi

\bibitem[{{Beck}(2000)}]{25}
{Beck}, J.~G. 2000, \solphys, 191, 47

\bibitem[{Belikov {et~al.}(2023)Belikov, Stark, Siegler, Por, Mennesson, Chen,
  Fogarty, Guyon, Juanola-Parramon, Krist, Mawet, Prada, Kasdin, Pueyo,
  Redmond, Ruane, Sirbu, Stapelfeldt, Trauger, \& Zimmerman}]{3}
Belikov, R., Stark, C., Siegler, N., {et~al.} 2023, in Techniques and
  Instrumentation for Detection of Exoplanets XI, ed. G.~J. Ruane, Vol. 12680,
  International Society for Optics and Photonics (SPIE), 126802G

\bibitem[{{Berdyugina}(2005)}]{22}
{Berdyugina}, S.~V. 2005, Living Reviews in Solar Physics, 2, 8

\bibitem[{Berk {et~al.}(1989)Berk, Bernstein, Robertson, {et~al.}}]{24}
Berk, A., Bernstein, L.~S., Robertson, D.~C., {et~al.} 1989, MODTRAN: A
  moderate resolution model for LOWTRAN 7

\bibitem[{Beuzit {et~al.}(2008)Beuzit, Feldt, Dohlen, Mouillet, Puget, Wildi,
  Abe, Antichi, Baruffolo, Baudoz, {et~al.}}]{a12}
Beuzit, J.-L., Feldt, M., Dohlen, K., {et~al.} 2008, in Ground-based and
  airborne instrumentation for astronomy II, Vol. 7014, SPIE, 476--487

\bibitem[{{Bowens} {et~al.}(2021){Bowens}, {Meyer, M. R.}, {Delacroix, C.},
  {Absil, O.}, {van Boekel, R.}, {Quanz, S. P.}, {Shinde, M.}, {Kenworthy, M.},
  {Carlomagno, B.}, {Orban de Xivry, G.}, {Cantalloube, F.}, \& {Pathak,
  P.}}]{R0}
{Bowens}, {Meyer, M. R.}, {Delacroix, C.}, {et~al.} 2021, A\&A, 653, A8

\bibitem[{{Bracewell}(1978)}]{4}
{Bracewell}, R.~N. 1978, \nat, 274, 780

\bibitem[{Chen {et~al.}(2021)Chen, Zhan, Youngblood, Wolf, Feinstein, \&
  Horton}]{392-2}
Chen, H., Zhan, Z., Youngblood, A., {et~al.} 2021, Nature Astronomy, 5, 298

\bibitem[{{Dannert, Felix A.} {et~al.}(2022){Dannert, Felix A.}, {Ottiger,
  Maurice}, {Quanz, Sascha P.}, {Laugier, Romain}, {Fontanet, Emile},
  {Gheorghe, Adrian}, {Absil, Olivier}, {Dandumont, Colin}, {Defrère, Denis},
  {Gascón, Carlos}, {Glauser, Adrian M.}, {Kammerer, Jens}, {Lichtenberg,
  Tim}, {Linz, Hendrik}, {Loicq, Jerôme}, \& {the LIFE collaboration}}]{9}
{Dannert, Felix A.}, {Ottiger, Maurice}, {Quanz, Sascha P.}, {et~al.} 2022,
  A\&A, 664, A22

\bibitem[{{Defr{\`e}re} {et~al.}(2024){Defr{\`e}re}, {Laugier}, {Martinod},
  {Garreau}, {Missiaen}, {Salman}, {Raskin}, {Dandumont}, {Ertel}, {Ireland},
  {Kraus}, {Labadie}, {Mazzolli}, {Medgyesi}, {Sanny}, {Absil},
  {{\'A}br{\'a}ham}, {Berger}, {Bonduelle}, {Bigioli}, {Bouzerand}, {Carter},
  {Cvetojevic}, {Courtney-Barrer}, {Glauser}, {Gross}, {Haubois}, {James},
  {Joo}, {Lagarde}, {L{\'e}ger}, {Leisenring}, {Loicq}, {Martin}, {Martinache},
  {Mezo}, {Morel}, {Morren}, {Ollivier}, {Robertson}, {Rousseau}, {Schofield},
  {Schuhler}, {Taras}, {Vandenbussche}, \& {Woillez}}]{R2}
{Defr{\`e}re}, D., {Laugier}, R., {Martinod}, M.-A., {et~al.} 2024, in Society
  of Photo-Optical Instrumentation Engineers (SPIE) Conference Series, Vol.
  13095, Optical and Infrared Interferometry and Imaging IX, ed. J.~{Kammerer},
  S.~{Sallum}, \& J.~{Sanchez-Bermudez}, 130950F

\bibitem[{{Defrère, D.} {et~al.}(2010){Defrère, D.}, {Absil, O.}, {Den
  Hartog, R.}, {Hanot, C.}, \& {Stark, C.}}]{R4}
{Defrère, D.}, {Absil, O.}, {Den Hartog, R.}, {Hanot, C.}, \& {Stark, C.}
  2010, A\&A, 509, A9

\bibitem[{Delorme {et~al.}(2021)Delorme, Jovanovic, Echeverri, Mawet,
  Kent~Wallace, Bartos, Cetre, Wizinowich, Ragland, Lilley, {et~al.}}]{a11}
Delorme, J.-R., Jovanovic, N., Echeverri, D., {et~al.} 2021, Journal of
  Astronomical Telescopes, Instruments, and Systems, 7, 035006

\bibitem[{{Ertel} {et~al.}(2020){Ertel}, {Defr{\`e}re}, {Hinz}, {Mennesson},
  {Kennedy}, {Danchi}, {Gelino}, {Hill}, {Hoffmann}, {Mazoyer}, {Rieke},
  {Shannon}, {Stapelfeldt}, {Spalding}, {Stone}, {Vaz}, {Weinberger},
  {Willems}, {Absil}, {Arbo}, {Bailey}, {Beichman}, {Bryden}, {Downey},
  {Durney}, {Esposito}, {Gaspar}, {Grenz}, {Haniff}, {Leisenring}, {Marion},
  {McMahon}, {Millan-Gabet}, {Montoya}, {Morzinski}, {Perera}, {Pinna}, {Pott},
  {Power}, {Puglisi}, {Roberge}, {Serabyn}, {Skemer}, {Su}, {Vaitheeswaran}, \&
  {Wyatt}}]{R5}
{Ertel}, S., {Defr{\`e}re}, D., {Hinz}, P., {et~al.} 2020, \aj, 159, 177

\bibitem[{{Fridlund} {et~al.}(2003){Fridlund}, {Henning}, \& {Lacoste}}]{6}
{Fridlund}, M., {Henning}, T., \& {Lacoste}, H., eds. 2003, ESA Special
  Publication, Vol. 539, {Towards other earths: DARWIN/TPF and the search for
  extrasolar terrestrial planets}

\bibitem[{{Froning} {et~al.}(2019){Froning}, {Kowalski}, {France}, {Loyd},
  {Schneider}, {Youngblood}, {Wilson}, {Brown}, {Berta-Thompson}, {Pineda},
  {Linsky}, {Rugheimer}, \& {Miguel}}]{21}
{Froning}, C.~S., {Kowalski}, A., {France}, K., {et~al.} 2019, \apjl, 871, L26

\bibitem[{Gao {et~al.}(2022)Gao, Liu, Yang, \& Zhou}]{a2}
Gao, D.-Y., Liu, H.-G., Yang, M., \& Zhou, J.-L. 2022, The Astronomical
  Journal, 164, 213

\bibitem[{{Gillon} {et~al.}(2017){Gillon}, {Triaud}, {Demory}, {Jehin}, {Agol},
  {Deck}, {Lederer}, {de Wit}, {Burdanov}, {Ingalls}, {Bolmont}, {Leconte},
  {Raymond}, {Selsis}, {Turbet}, {Barkaoui}, {Burgasser}, {Burleigh}, {Carey},
  {Chaushev}, {Copperwheat}, {Delrez}, {Fernandes}, {Holdsworth}, {Kotze}, {Van
  Grootel}, {Almleaky}, {Benkhaldoun}, {Magain}, \& {Queloz}}]{12}
{Gillon}, M., {Triaud}, A. H.~M.~J., {Demory}, B.-O., {et~al.} 2017, \nat, 542,
  456

\bibitem[{Golovin {et~al.}(2023)Golovin, Reffert, Just, Jordan, Vani, \&
  Jahrei{\ss}}]{248}
Golovin, A., Reffert, S., Just, A., {et~al.} 2023, Astronomy \& Astrophysics,
  670, A19

\bibitem[{Gopalswamy {et~al.}(2003)Gopalswamy, Lara, Yashiro, Nunes, \&
  Howard}]{388}
Gopalswamy, N., Lara, A., Yashiro, S., Nunes, S., \& Howard, R.~A. 2003, Solar
  Variability as an Input to the Earth's Environment, 535, 403

\bibitem[{{Hansen, Jonah T.} {et~al.}(2022){Hansen, Jonah T.}, {Ireland,
  Michael J.}, \& {the LIFE Collaboration}}]{8}
{Hansen, Jonah T.}, {Ireland, Michael J.}, \& {the LIFE Collaboration}. 2022,
  A\&A, 664, A52

\bibitem[{{Hauschildt} {et~al.}(1999{\natexlab{a}}){Hauschildt}, {Allard}, \&
  {Baron}}]{17}
{Hauschildt}, P.~H., {Allard}, F., \& {Baron}, E. 1999{\natexlab{a}}, \apj,
  512, 377

\bibitem[{{Hauschildt} {et~al.}(1999{\natexlab{b}}){Hauschildt}, {Allard},
  {Ferguson}, {Baron}, \& {Alexander}}]{16}
{Hauschildt}, P.~H., {Allard}, F., {Ferguson}, J., {Baron}, E., \& {Alexander},
  D.~R. 1999{\natexlab{b}}, \apj, 525, 871

\bibitem[{{Henry} \& {RECONS Team}(2002)}]{23}
{Henry}, T.~J. \& {RECONS Team}. 2002, in American Astronomical Society Meeting
  Abstracts, Vol. 201, American Astronomical Society Meeting Abstracts, 104.05

\bibitem[{Howard {et~al.}(2018)Howard, Tilley, Corbett, Youngblood, Loyd,
  Ratzloff, Law, Fors, Del~Ser, Shkolnik, {et~al.}}]{a7}
Howard, W.~S., Tilley, M.~A., Corbett, H., {et~al.} 2018, The Astrophysical
  Journal Letters, 860, L30

\bibitem[{Jackman {et~al.}(2023)Jackman, Shkolnik, Million, Fleming,
  Richey-Yowell, \& Loyd}]{a6}
Jackman, J.~A., Shkolnik, E.~L., Million, C., {et~al.} 2023, Monthly Notices of
  the Royal Astronomical Society, 519, 3564

\bibitem[{Kervella {et~al.}(2017)Kervella, Th{\'e}venin, \& Lovis}]{a1}
Kervella, P., Th{\'e}venin, F., \& Lovis, C. 2017, Astronomy \& Astrophysics,
  598, L7

\bibitem[{{Lay}(2006)}]{15}
{Lay}, O.~P. 2006, in Society of Photo-Optical Instrumentation Engineers (SPIE)
  Conference Series, Vol. 6268, Advances in Stellar Interferometry, ed. J.~D.
  {Monnier}, M.~{Sch{\"o}ller}, \& W.~C. {Danchi}, 62681A

\bibitem[{{Lovis} \& {Fischer}(2010)}]{13}
{Lovis}, C. \& {Fischer}, D. 2010, in Exoplanets, ed. S.~{Seager}, 27--53

\bibitem[{{Loyd} {et~al.}(2019){Loyd}, {Shkolnik}, {Barman}, {Peacock},
  {Schneider}, {Meadows}, \& {Pagano}}]{20}
{Loyd}, P., {Shkolnik}, E.~L., {Barman}, T., {et~al.} 2019, in American
  Astronomical Society Meeting Abstracts, Vol. 233, American Astronomical
  Society Meeting Abstracts \#233, 204.04

\bibitem[{Macintosh {et~al.}(2014)Macintosh, Graham, Ingraham, Konopacky,
  Marois, Perrin, Poyneer, Bauman, Barman, Burrows, {et~al.}}]{a13}
Macintosh, B., Graham, J.~R., Ingraham, P., {et~al.} 2014, proceedings of the
  National Academy of Sciences, 111, 12661

\bibitem[{{Martinache, Frantz} \& {Ireland, Michael J.}(2018)}]{5}
{Martinache, Frantz} \& {Ireland, Michael J.} 2018, A\&A, 619, A87

\bibitem[{{McIntosh} {et~al.}(2014){McIntosh}, {Wang}, {Leamon}, {Davey},
  {Howe}, {Krista}, {Malanushenko}, {Markel}, {Cirtain}, {Gurman}, {Pesnell},
  \& {Thompson}}]{421}
{McIntosh}, S.~W., {Wang}, X., {Leamon}, R.~J., {et~al.} 2014, \apj, 792, 12

\bibitem[{{Miles} {et~al.}(2023){Miles}, {Biller}, {Patapis}, {Worthen},
  {Rickman}, {Hoch}, {Skemer}, {Perrin}, {Whiteford}, {Chen}, {Sargent},
  {Mukherjee}, {Morley}, {Moran}, {Bonnefoy}, {Petrus}, {Carter}, {Choquet},
  {Hinkley}, {Ward-Duong}, {Leisenring}, {Millar-Blanchaer}, {Pueyo}, {Ray},
  {Sallum}, {Stapelfeldt}, {Stone}, {Wang}, {Absil}, {Balmer}, {Boccaletti},
  {Bonavita}, {Booth}, {Bowler}, {Chauvin}, {Christiaens}, {Currie},
  {Danielski}, {Fortney}, {Girard}, {Grady}, {Greenbaum}, {Henning}, {Hines},
  {Janson}, {Kalas}, {Kammerer}, {Kennedy}, {Kenworthy}, {Kervella}, {Lagage},
  {Lew}, {Liu}, {Macintosh}, {Marino}, {Marley}, {Marois}, {Matthews},
  {Matthews}, {Mawet}, {McElwain}, {Metchev}, {Meyer}, {Molliere}, {Pantin},
  {Quirrenbach}, {Rebollido}, {Ren}, {Schneider}, {Vasist}, {Wyatt}, {Zhou},
  {Briesemeister}, {Bryan}, {Calissendorff}, {Cantalloube}, {Cugno}, {De
  Furio}, {Dupuy}, {Factor}, {Faherty}, {Fitzgerald}, {Franson}, {Gonzales},
  {Hood}, {Howe}, {Kraus}, {Kuzuhara}, {Lagrange}, {Lawson}, {Lazzoni}, {Liu},
  {Llop-Sayson}, {Lloyd}, {Martinez}, {Mazoyer}, {Quanz}, {Redai}, {Samland},
  {Schlieder}, {Tamura}, {Tan}, {Uyama}, {Vigan}, {Vos}, {Wagner}, {Wolff},
  {Ygouf}, {Zhang}, {Zhang}, \& {Zhang}}]{1}
{Miles}, B.~E., {Biller}, B.~A., {Patapis}, P., {et~al.} 2023, \apjl, 946, L6

\bibitem[{{Morris} {et~al.}(2018){Morris}, {Agol}, {Davenport}, \&
  {Hawley}}]{11}
{Morris}, B.~M., {Agol}, E., {Davenport}, J. R.~A., \& {Hawley}, S.~L. 2018,
  \apj, 857, 39

\bibitem[{O'Neal {et~al.}(1998)O'Neal, Neff, \& Saar}]{a4}
O'Neal, D., Neff, J.~E., \& Saar, S.~H. 1998, The Astrophysical Journal, 507,
  919

\bibitem[{{Oshagh} {et~al.}(2013){Oshagh}, {Santos}, {Boisse}, {Bou{\'e}},
  {Montalto}, {Dumusque}, \& {Haghighipour}}]{10}
{Oshagh}, M., {Santos}, N.~C., {Boisse}, I., {et~al.} 2013, \aap, 556, A19

\bibitem[{Qiao-Yang {et~al.}(2023)Qiao-Yang, Shen-Wei, \& Hui-Gen}]{a8}
Qiao-Yang, H., Shen-Wei, Z., \& Hui-Gen, L. 2023, Publications of the
  Astronomical Society of the Pacific, 135, 094401

\bibitem[{{Quanz, S. P.} {et~al.}(2022){Quanz, S. P.}, {Ottiger, M.},
  {Fontanet, E.}, {Kammerer, J.}, {Menti, F.}, {Dannert, F.}, {Gheorghe, A.},
  {Absil, O.}, {Airapetian, V. S.}, {Alei, E.}, {Allart, R.}, {Angerhausen,
  D.}, {Blumenthal, S.}, {Buchhave, L. A.}, {Cabrera, J.}, {Carrión-González,
  Ó.}, {Chauvin, G.}, {Danchi, W. C.}, {Dandumont, C.}, {Defrére, D.}, {Dorn,
  C.}, {Ehrenreich, D.}, {Ertel, S.}, {Fridlund, M.}, {García Muñoz, A.},
  {Gascón, C.}, {Girard, J. H.}, {Glauser, A.}, {Grenfell, J. L.}, {Guidi,
  G.}, {Hagelberg, J.}, {Helled, R.}, {Ireland, M. J.}, {Janson, M.},
  {Kopparapu, R. K.}, {Korth, J.}, {Kozakis, T.}, {Kraus, S.}, {Léger, A.},
  {Leedjärv, L.}, {Lichtenberg, T.}, {Lillo-Box, J.}, {Linz, H.}, {Liseau,
  R.}, {Loicq, J.}, {Mahendra, V.}, {Malbet, F.}, {Mathew, J.}, {Mennesson,
  B.}, {Meyer, M. R.}, {Mishra, L.}, {Molaverdikhani, K.}, {Noack, L.}, {Oza,
  A. V.}, {Pallé, E.}, {Parviainen, H.}, {Quirrenbach, A.}, {Rauer, H.},
  {Ribas, I.}, {Rice, M.}, {Romagnolo, A.}, {Rugheimer, S.}, {Schwieterman, E.
  W.}, {Serabyn, E.}, {Sharma, S.}, {Stassun, K. G.}, {Szulágyi, J.}, {Wang,
  H. S.}, {Wunderlich, F.}, {Wyatt, M. C.}, \& {the LIFE Collaboration}}]{7}
{Quanz, S. P.}, {Ottiger, M.}, {Fontanet, E.}, {et~al.} 2022, A\&A, 664, A21

\bibitem[{{Queloz} {et~al.}(2001){Queloz}, {Henry}, {Sivan}, {Baliunas},
  {Beuzit}, {Donahue}, {Mayor}, {Naef}, {Perrier}, \& {Udry}}]{14}
{Queloz}, D., {Henry}, G.~W., {Sivan}, J.~P., {et~al.} 2001, \aap, 379, 279

\bibitem[{Sabotta {et~al.}(2021)Sabotta, Schlecker, Chaturvedi, Guenther,
  Rodr{\'\i}guez, S{\'a}nchez, Caballero, Shan, Reffert, Ribas, {et~al.}}]{249}
Sabotta, S., Schlecker, M., Chaturvedi, P., {et~al.} 2021, Astronomy \&
  Astrophysics, 653, A114

\bibitem[{Segura {et~al.}(2010)Segura, Walkowicz, Meadows, Kasting, \&
  Hawley}]{392-1}
Segura, A., Walkowicz, L.~M., Meadows, V., Kasting, J., \& Hawley, S. 2010,
  Astrobiology, 10, 751

\bibitem[{{Serabyn} {et~al.}(2012){Serabyn}, {Mennesson}, {Colavita},
  {Koresko}, \& {Kuchner}}]{R1}
{Serabyn}, E., {Mennesson}, B., {Colavita}, M.~M., {Koresko}, C., \& {Kuchner},
  M.~J. 2012, \apj, 748, 55

\bibitem[{Shulyak {et~al.}(2017)Shulyak, Reiners, Engeln, Malo, Yadav, Morin,
  \& Kochukhov}]{a3}
Shulyak, D., Reiners, A., Engeln, A., {et~al.} 2017, Nature Astronomy, 1, 0184

\bibitem[{Solanki(2003)}]{18}
Solanki, S.~K. 2003, The Astronomy and Astrophysics Review, 11, 153

\bibitem[{Vasquez {et~al.}(2013)Vasquez, Schreier, Garc{\'\i}a, Kitzmann,
  Patzer, Rauer, \& Trautmann}]{a9}
Vasquez, M., Schreier, F., Garc{\'\i}a, S.~G., {et~al.} 2013, Astronomy \&
  Astrophysics, 549, A26

\bibitem[{Zhao {et~al.}(2024)Zhao, Hua, Cheng, Li, \& Ding}]{389}
Zhao, Z., Hua, Z., Cheng, X., Li, Z., \& Ding, M. 2024, The Astrophysical
  Journal, 961, 130

\bibitem[{{Zhilyaev, B. E.} {et~al.}(2007){Zhilyaev, B. E.}, {Romanyuk, Ya.
  O.}, {Svyatogorov, O. A.}, {Verlyuk, I. A.}, {Kaminsky, B.}, {Andreev, M.},
  {Sergeev, A. V.}, {Gershberg, R. E.}, {Lovkaya, M. N.}, {Avgoloupis, S. J.},
  {Seiradakis, J. H.}, {Contadakis, M. E.}, {Antov, A. P.},
  {Konstantinova-Antova, R. K.}, \& {Bogdanovski, R.}}]{19}
{Zhilyaev, B. E.}, {Romanyuk, Ya. O.}, {Svyatogorov, O. A.}, {et~al.} 2007,
  A\&A, 465, 235

\bibitem[{Zieba {et~al.}(2023)Zieba, Kreidberg, Ducrot, Gillon, Morley,
  Schaefer, Tamburo, Koll, Lyu, Acu{\~n}a, {et~al.}}]{a10}
Zieba, S., Kreidberg, L., Ducrot, E., {et~al.} 2023, Nature, 620, 746

\bibitem[{{Zurlo, A.} {et~al.}(2016){Zurlo, A.}, {Vigan, A.}, {Galicher, R.},
  {Maire, A.-L.}, {Mesa, D.}, {Gratton, R.}, {Chauvin, G.}, {Kasper, M.},
  {Moutou, C.}, {Bonnefoy, M.}, {Desidera, S.}, {Abe, L.}, {Apai, D.},
  {Baruffolo, A.}, {Baudoz, P.}, {Baudrand, J.}, {Beuzit, J.-L.}, {Blancard,
  P.}, {Boccaletti, A.}, {Cantalloube, F.}, {Carle, M.}, {Cascone, E.},
  {Charton, J.}, {Claudi, R. U.}, {Costille, A.}, {de Caprio, V.}, {Dohlen,
  K.}, {Dominik, C.}, {Fantinel, D.}, {Feautrier, P.}, {Feldt, M.}, {Fusco,
  T.}, {Gigan, P.}, {Girard, J. H.}, {Gisler, D.}, {Gluck, L.}, {Gry, C.},
  {Henning, T.}, {Hugot, E.}, {Janson, M.}, {Jaquet, M.}, {Lagrange, A.-M.},
  {Langlois, M.}, {Llored, M.}, {Madec, F.}, {Magnard, Y.}, {Martinez, P.},
  {Maurel, D.}, {Mawet, D.}, {Meyer, M. R.}, {Milli, J.}, {Moeller-Nilsson,
  O.}, {Mouillet, D.}, {Origné, A.}, {Pavlov, A.}, {Petit, C.}, {Puget, P.},
  {Quanz, S. P.}, {Rabou, P.}, {Ramos, J.}, {Rousset, G.}, {Roux, A.},
  {Salasnich, B.}, {Salter, G.}, {Sauvage, J.-F.}, {Schmid, H. M.}, {Soenke,
  C.}, {Stadler, E.}, {Suarez, M.}, {Turatto, M.}, {Udry, S.}, {Vakili, F.},
  {Wahhaj, Z.}, {Wildi, F.}, \& {Antichi, J.}}]{2}
{Zurlo, A.}, {Vigan, A.}, {Galicher, R.}, {et~al.} 2016, A\&A, 587, A57

\end{thebibliography}
\bibliographystyle{aa}

\end{document}